\documentclass[runningheads]{llncs}
\usepackage{amsmath}
\usepackage{amssymb}
\usepackage{booktabs}
\usepackage{epsfig}
\usepackage{graphicx}
\usepackage{url}
\usepackage{bm}
\usepackage{tikz}
\usepackage{comment}
\usepackage{amsmath,amssymb} 
\usepackage{color}
\usepackage{newtxmath}
\DeclareUnicodeCharacter{2212}{-}
\usepackage[pagebackref=true,breaklinks=true,letterpaper=true,colorlinks,bookmarks=false]{hyperref}
\usepackage[accsupp]{axessibility}  
\usepackage{multirow}
\usepackage{booktabs}

\usepackage{subcaption}

\renewcommand{\thefootnote}{*}
\usepackage{newtxmath}

\begin{document}
\pagestyle{headings}
\mainmatter
\def\ECCVSubNumber{4402}  

\title{Optimizing Image Compression via Joint Learning with Denoising} 

\titlerunning{Optimizing Image Compression via Joint Learning with Denoising}
%
\author{Ka Leong Cheng\inst{\star}\index{Cheng, Ka Leong} \and
Yueqi Xie\inst{\star} \and
Qifeng Chen}
\authorrunning{K. L. Cheng et al.}
%
\institute{The Hong Kong Univeristy of Science and Technology, Hong Kong, China\\
\email{\{klchengad,yxieay\}@connect.ust.hk, cqf@ust.hk}}
\maketitle

\renewcommand{\thefootnote}{\fnsymbol{footnote}}
\footnotetext[1]{Joint first authors}

\begin{abstract}
High levels of noise usually exist in today's captured images due to the relatively small sensors equipped in the smartphone cameras, where the noise brings extra challenges to lossy image compression algorithms. Without the capacity to tell the difference between image details and noise, general image compression methods allocate additional bits to explicitly store the undesired image noise during compression and restore the unpleasant noisy image during decompression. Based on the observations, we optimize the image compression algorithm to be noise-aware as joint denoising and compression to resolve the bits misallocation problem. The key is to transform the original noisy images to noise-free bits by eliminating the undesired noise during compression, where the bits are later decompressed as clean images. Specifically, we propose a novel two-branch, weight-sharing architecture with plug-in feature denoisers to allow a simple and effective realization of the goal with little computational cost. Experimental results show that our method gains a significant improvement over the existing baseline methods on both the synthetic and real-world datasets. Our source code is available at:~\url{https://github.com/felixcheng97/DenoiseCompression}.

\end{abstract}

\section{Introduction}
\label{sec:intro}
Lossy image compression has been studied for decades with essential applications in media storage and transmission. Many traditional algorithms~\cite{rabbani2002jpeg2000,wallace1992jpeg} and learned methods~\cite{balle2016end,balle2018variational,cheng2020learned,minnen2018joint,xie2021enhanced} are proposed and widely used. Thanks to the fast development of mobile devices, smartphones are becoming the most prevalent and convenient choice of photography for sharing. However, the captured images usually contain high levels of noise due to the limited sensor and aperture size in smartphone cameras~\cite{abdelhamed2018sidd}. Since existing compression approaches are designed for general images, the compressors treat the noise as ``crucial'' information and explicitly allocate bits to store it, even though noise is usually undesired for common users. The image noise can further degrade the compression quality, especially at medium and high bit rates~\cite{alshaykh1998lossy,ponomarenko2010noisy}. Concerning these aspects, we see the crucial need for an image compression method with the capacity of noise removal during the compression process.

A natural and straightforward solution is to go through a sequential pipeline of individual denoising and compression methods. However, a simple combination of separate models can be sub-optimal for this joint task. On the one hand, sequential methods introduce additional time overhead due to the intermediate results, leading to a lower efficiency than a united solution. The inferior efficiency can limit their practical applications, especially on mobile devices. On the other hand, a sequential solution suffers from the accumulation of errors and information loss in the individual models. Most image denoising algorithms have strong capabilities of noise removal for the flat regions but somehow over-smooth the image details~\cite{xing2021end}. However, the details are the critical parts of information that need to be kept for compression. Lossy image compression algorithms save bits through compressing local patterns with a certain level of information loss, particularly for the high-frequency patterns. However, both image details and noise are considered high-frequency information, so a general image compressor is likely to eliminate some useful high-frequency details while misallocating bits to store the unwanted noise instead. In the area of image processing, many researchers explore to develop joint solutions instead of using sequential approaches, such as combined problems of joint image demosaicing, denoising, or super-resolution~\cite{gharbi2016deep,liu2020joint,xing2021end}.

In this paper, we contribute a joint method to optimize the image compression algorithm via joint learning with denoising. The key challenge of this joint task is to resolve the bit misallocation issue on the undesired noise when compressing the noisy images. In other words, the joint denoising and compression method needs to eliminate only the image noise while preserving the desired high-frequency content so that no extra bits are wastefully allocated for encoding the noise information in the images. Some existing works attempt to integrate the denoising problem into image compression algorithms. 
Prior works~\cite{gonzalez2018joint,preciozzi2017joint} focus on the decompression procedure and propose joint image denoise-decompression algorithms, which take the noisy wavelets coefficients as input to restore the clean images, but leave the compressing part untouched. A recent work~\cite{testolina2021towards} attempts to tackle this task by adding several convolutional layers into the decompressor to denoise the encoded latent features on the decoding side. However, their networks can inevitably use additional bits to store the noise in the latent features since there are no particular designs of modules or supervision for denoising in the compressor, leading to their inferior performance compared to the sequentially combined denoising and compression solutions. 

We design an end-to-end trainable network with a simple yet effective novel two-branch design (a denoising branch and a guidance branch) to resolve the bit misallocation problem in joint image denoising and compression. Specifically, we hope to pose explicit supervision on the encoded latent features to ensure it is noise-free so that we can eliminate high-frequency noise while, to a great extent, preserving useful information. During training, the denoising and guidance branches have shared encoding modules to obtain noisy features from the noisy input image and the guiding features from the clean input image, respectively; efficient denoising modules are plugged into the denoising branch to denoise the noisy features as noise-free latent codes. The explicit supervision is posed in high-dimensional space from the guiding features to the encoded latent codes. In this way, we can train the denoiser to help learn a noise-free representation. Note that the guidance branch is disabled during inference.

We conduct extensive experiments for joint image denoising and compression on both the synthetic data under various noise levels and the real-world SIDD~\cite{abdelhamed2018sidd}. Our main contributions are as follows:
\begin{itemize}
    \renewcommand{\labelitemi}{\textbullet}
    \item We optimize image compression on noisy images through joint learning with denoising, aiming to avoid bit misallocation for the undesired noise. Our method outperforms baseline methods on both the synthetic and real-world datasets by a large margin.
    \item We propose an end-to-end joint denoising and compression network with a novel two-branch design to explicitly supervise the network to eliminate noise while preserving high-frequency details in the compression process.
    \item Efficient plug-in feature denoisers are designed and incorporated into the denoising branch to enable the denoising capacity of the compressor with only little addition of complexity during inference time.
\end{itemize}

\section{Related Work}
\subsection{Image Denoising}
Image denoising is an age-long studied task with many traditional methods proposed over the past decades. They typically rely on certain pre-defined assumptions of noise distribution, including sparsity of image gradients~\cite{chambolle2004algorithm,rudin1992nonlinear} and similarity of image patches~\cite{dabov2007color,gu2014weighted}. With the rapid development of deep learning, some methods~\cite{chen2018learning,guan2019node,wang2020practical} utilize CNNs to improve the image denoising performance based on the synthetic~\cite{foi2008practical,wang2019enhancing,wei2020physics} and real-world datasets, including DND~\cite{plotz2017benchmarking}, SIDD~\cite{abdelhamed2018sidd}, and SID~\cite{chen2018learning}. Some works~\cite{guo2019toward,kim2020transfer,zhang2018ffdnet} focus on adapting solutions from synthetic datasets to real-world scenarios. Some current state-of-the-art methods are proposed to enhance performance further, including DANet~\cite{yue2020dual} utilizing an adversarial framework and InvDN~\cite{liu2021invertible} leveraging the invertible neural networks. However, many learning-based solutions rely on heavy denoising models, which are practically inefficient for the joint algorithms, especially in real-world applications.

\subsection{Lossy Image Compression}
Many traditional lossy image compression solutions~\cite{bpg,webp,vvc,rabbani2002jpeg2000,wallace1992jpeg} are widely proposed for practical usage. They map an image to quantized latent codes through hand-crafted transformations and compress them using entropy coding. With vast amounts of data available these days, many learning-based solutions are developed to learn a better transformation between image space and feature space. RNN-based methods~\cite{johnston2018improved,toderici2016variable,toderici2017full} are utilized to iteratively encode residual information in the images, while most of the recent solutions are based on variational autoencoders (VAEs)~\cite{balle2017end,theis2017lossy} to optimize the whole image compression process directly. Some methods~\cite{balle2018variational,cheng2020learned,guo2020context,hu2020coarse,lee2019context,minnen2018joint,minnen2020channel} focus on improving the entropy models to parameterize the latent code distribution more accurately. Some others~\cite{lin2020spatial,xie2021enhanced} design stronger architectures to learn better transformations for image compression. For example, Lin et al.~\cite{lin2020spatial} introduce spatial RNNs to reduce spatial redundancy, Mentzer et al.~\cite{mentzer2020high} integrate generative adversarial networks for high-fidelity generative image compression, and Xie et al.~\cite{xie2021enhanced} utilize invertible neural networks to form a better reversible process. However, these existing compression methods generally do not consider the image noise in their designs.

\subsection{Joint Solutions}
A series of operations are usually included in a whole image or video processing pipeline, while a pipeline with separate solutions can suffer from the accumulation of errors from individual methods. Thus, many joint solutions have been proposed for various combinations of tasks. Several widely-studied ones for image processing include joint denoising and demosaicing~\cite{condat2012joint,ehret2019joint,gharbi2016deep,khashabi2014joint,klatzer2016learning}, joint denoising and super-resolution~\cite{zhang2018learning}, and joint demosaicing and super-resolution~\cite{farsiu2004multiframe,vandewalle2007joint,xu2020joint}. Recently, Xing et al.~\cite{xing2021end} further propose to solve a joint triplet problem of image denoising, demosaicing, and super-resolution. As for video processing, Norkin et al.~\cite{norkin2018film} integrate the idea of separating film grain from video content into the AV1 video codec, which can be viewed as a joint solution of video denoising and compression. However, the task for joint image denoising and compression has not received much attention yet important. Some works like~\cite{gonzalez2018joint,preciozzi2017joint} only target at building decompression methods that restore clean images from the noisy bits by integrating the denoising procedure in the decompression process. A recent work~\cite{testolina2021towards} also incorporates the denoising idea into image decompression by performing denoising on the encoded latent codes during decompression. These approaches cannot achieve our goal of solving the bits misallocation problem and cannot achieve pleasant rate-distortion performance since denoising during decompression cannot produce noise-free bits.

\section{Problem Specification} 
We wish to build an image compression method that takes noise removal into consideration during compression since noise is usually unwanted for general users while requiring additional bits for storage. Hence, the benefit of such a compressor lies in saving storage for the unwanted noise during the compression process.
Formally, given a noisy image $\mathbf{\tilde{x}}$ with its corresponding clean ground truth image $\mathbf{x}$, the compressor takes $\mathbf{\tilde{x}}$ as input to denoise and compress it into denoised bitstreams. We can later decompress the bitstreams to get the denoised image $\mathbf{\hat{x}}$. Meanwhile, instead of sequentially doing denoising and successive compression or vice versa, we require the whole process to be end-to-end optimized as a united system to improve efficiency and avoid accumulation of errors.

\subsection{Selection of Datasets}
It is desirable to have a large number of diverse samples to train a well-performing learned image compression network. Many real-world datasets such as DND~\cite{plotz2017benchmarking}, SIDD~\cite{abdelhamed2018sidd}, and SID~\cite{chen2018learning} have been proposed for image denoising with noisy-clean image pairs. However, they generally have limited training samples, scene diversities, or noise levels because collecting a rich (large scale, various gains, illuminance, etc.) real-world dataset typically includes much time-consuming labor work. In contrast, synthetic data is cheap and unlimited, and it is flexible to synthesize images with different levels of noise. Therefore, the main experiments in this paper are carried out using synthetic data; additional experiments on the SIDD~\cite{abdelhamed2018sidd} are also conducted to further verify the effectiveness of our method. We use the SIDD only for real-world datasets because RNI15~\cite{lebrun2015the} has no clean ground truths; SID~\cite{chen2018learning} is for image denoising in the dark; DND~\cite{plotz2017benchmarking} only allows 5 monthly submissions, which is not suitable for the image compression task that requires evaluations at various bit rates.

\subsection{Noise Synthesis}
We use a similar strategy as in~\cite{mildenhall2018kpn} to do noise synthesis in raw, where the following sRGB gamma correction function $\Gamma$ is used to transform between the image sRGB domain $\mathbf{X}$ and the raw linear domain $\mathbf{Y}$:
\begin{equation}
\mathbf{X}=\Gamma(\mathbf{Y})=\left\{
    \begin{array}{lr}
        m \mathbf{Y}, & \mathbf{Y} \leq b, \\
        (1 + a) \mathbf{Y}^{1/\gamma} - a, & \mathbf{Y} > b,
    \end{array}
\right.
\end{equation}
where $a = 0.055$, $b = 0.0031308$, $m = 12.92$, $\gamma = 2.4$.
Specifically, the inverse function $\Gamma^{-1}$ is first applied on sRGB image $\mathbf{x}$ to get the raw image $\mathbf{y} = \Gamma^{-1}(\mathbf{x})$. The noise in raw images is defined as the standard deviation of the linear signal, ranging from $0$ to $1$. Given a true signal intensity $y_p$ at position $p$, the corresponding noisy measurement $\tilde{y}_p$ in the noisy raw image $\mathbf{\tilde{y}}$ is estimated by a two-parameter signal-dependent Gaussian distribution~\cite{healey1994radiometric}:
\begin{equation}
    \tilde{y}_p \sim \mathcal{N} (y_p, \sigma_s y_p + \sigma_r^2),
\end{equation}
where $\sigma_s$ and $\sigma_r$ denote the shot and readout noise parameters, respectively, indicating different sensor gains (ISO). After the noise synthesis in raw space, we obtain our noisy sRGB image $\mathbf{\tilde{x}} = \Gamma(\mathbf{\tilde{y}})$.

\begin{figure}[t]
\begin{center}
\includegraphics[width=0.99\linewidth]{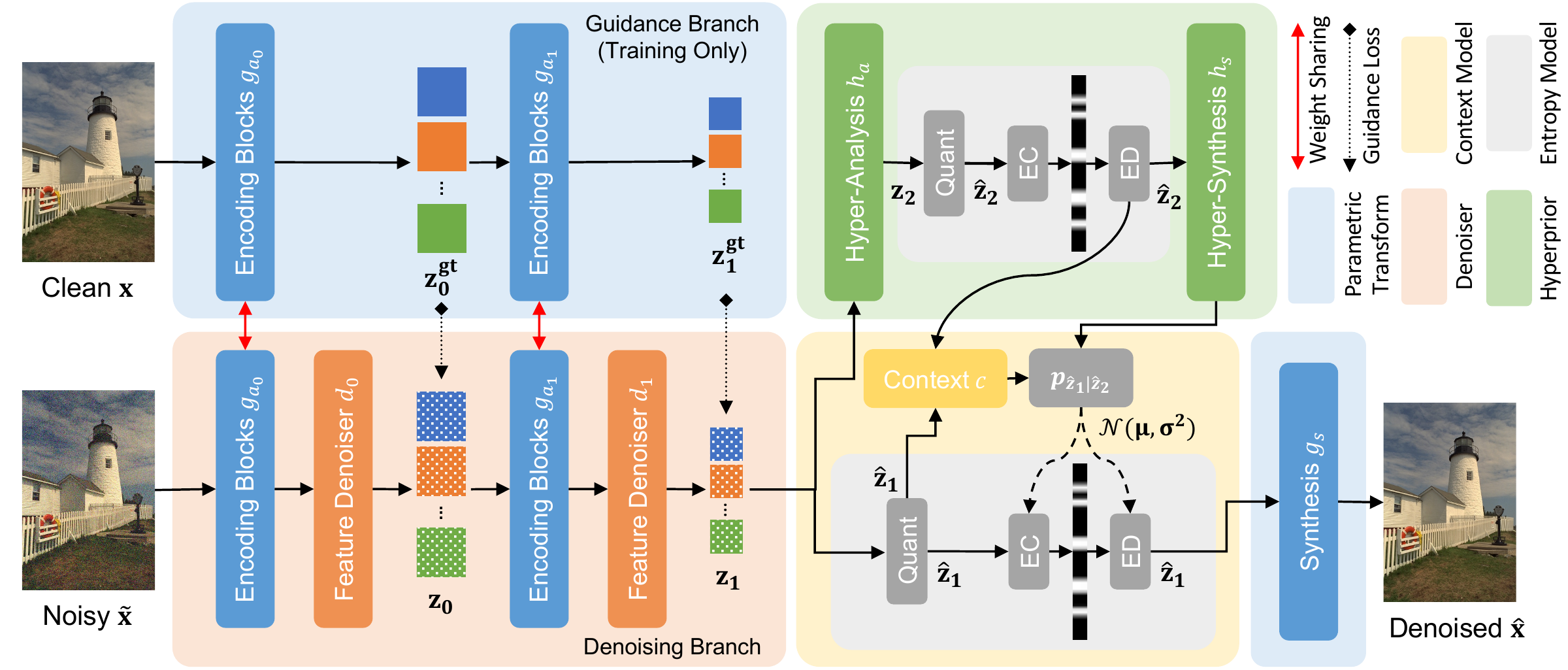}
\end{center}
\caption{Overview of the two-branch design of our proposed network, which is first pre-trained on clean images and successively fine-tuned on noisy-clean image pairs. In the top left of the figure, the clean image goes through the guidance branch for the two-level guiding features; in the bottom left, the noisy image is fed into the denoising branch to obtain the two-level denoised features. Note that the guidance branch is for training only, and that the denoising branch (orange part) and the denoisers (orange blocks) are only activated during fine-tuning and used for inference. The right half of the figure contains the common hyperprior, entropy models, context model, and synthesis transform used in the recent learned compression methods~\cite{cheng2020learned,minnen2018joint}.}
\label{fig:overview}
\end{figure}

\section{Method}
\label{sec:method}
Our joint denoising and compression method is inherently an image compression algorithm with the additional capacity to remove undesirable noise. Hence, the proposed method for image denoise-compression is built upon the learned image compression methods. Fig.~\ref{fig:overview} shows an overview of the proposed method. Our network contains a novel two-branch design for the training process, where the guiding features in the guidance branch pose explicit supervision on the denoised features in the denoising branch during the compression process.

\subsection{Network Design}
\textbf{Overall workflow.} The network contains a parametric analysis transform $g_a$ (containing $g_{a_0}$ and $g_{a_1}$) with plug-in feature denoisers $d$ (containing $d_0$ and $d_1$) to encode and denoise the input noisy image $\mathbf{\tilde{x}}$ into some denoised latent features $\mathbf{z_1}$. Discrete quantization is then applied to obtain the quantized latent features $\mathbf{\hat{z}_1}$. Instead of using the non-differentiable discrete rounding function during training, we add a uniform noise $\mathcal{U}(-0.5, 0.5)$ on top of $\mathbf{z_1}$ to get $\mathbf{\tilde{z}_1}$, which can be view as an approximation of the discrete quantization process~\cite{balle2017end}. For notation simplicity, we use $\mathbf{\hat{z}_1}$ to represent both $\mathbf{\hat{z}_1}$ and $\mathbf{\tilde{z}_1}$ in this paper. Then accordingly, we have a parametric synthesis transform $g_s$ that decodes $\mathbf{\hat{z}_1}$ to obtain the denoised image $\mathbf{\hat{x}}$. The parametric transforms $g_a$, $g_s$ and the denoiser $d$ formulate a basic variational model for the joint image denoising and compression task.

As discussed in Ball\'{e} et al.~\cite{balle2018variational}, there still remain significant spatial dependencies within the latent features $\mathbf{\hat{z}_1}$ using a basic variational model. Hence, a similar scale hyperprior is appended on top of the basic variational model. In particular, the hyperprior contains parametric transform $h_a$ to model the spatial dependencies and obtain the additional latent features $\mathbf{z_2}$, so that we can assume that the target variables $\mathbf{z_1}$ conditioned on the introduced latent $\mathbf{z_1}$ are independent~\cite{bishop1998latent}. We adopt the same uniform noise strategy on $\mathbf{z_2}$ to obtain $\mathbf{\tilde{z}_2}$ during training and perform discrete quantization for $\mathbf{\hat{z}_2}$ during testing. Similarly, we use $\mathbf{\hat{z}_2}$ to represent $\mathbf{\hat{z}_2}$ and $\mathbf{\tilde{z}_2}$ for notation simplicity. Together with the causal context model $c$, another parametric synthesis transform $h_s$ transforms $\mathbf{\hat{z}_2}$ to estimate the means $\hat{\bm{\muup}}$ and standard deviations $\hat{\bm{\sigmaup}}$ for the latent features $\mathbf{\hat{z}_1}$ so that each element of the latent features is modeled as mean and scale Gaussian~\cite{minnen2018joint}:
\begin{equation}
    p_{\mathbf{\hat{z}_1}|\mathbf{\hat{z}_2}} \sim \mathcal{N}(\hat{\bm{\muup}}, \hat{\bm{\sigmaup}}^2).
\end{equation}
Similar to~\cite{balle2017end}, the distribution of $\mathbf{\hat{z}_2}$ is modeled as $p_{\mathbf{\hat{z}_2}|\theta}$ by a non-parametric, factorized entropy model $\theta$ because the prior knowledge is not available for $\mathbf{\hat{z}_2}$.

\textbf{Two-branch architecture.}
The noisy image $\mathbf{\tilde{x}}$ and the corresponding clean image $\mathbf{x}$ are fed into the denoising and guidance branches, respectively. Similar to many denoising methods~\cite{chang2020spatial,cheng2021nbnet}, the plug-in denoisers $d_0$ and $d_1$ are designed in a multiscale manner. The two-level guiding features $\mathbf{z_0^{gt}}$ and $\mathbf{z_1^{gt}}$ are obtained by the parametric analysis $g_{a_0}$ and $g_{a_1}$, respectively; the two-level denoised features $\mathbf{z_0}$ and $\mathbf{z_1}$ are obtained by the parametric analysis $g_{a_0}$ and $g_{a_1}$ plus the denoisers $d_0$ and $d_1$, respectively:
\begin{equation}
\begin{alignedat}{3}
    & \mathbf{z_0^{gt}} = g_{a_0} (\mathbf{x}),\quad        && \mathbf{z_0} = g_{a_0} (\mathbf{\tilde{x}}) && + d_0 (g_{a_0} (\mathbf{\tilde{x}})), \\
    & \mathbf{z_1^{gt}} = g_{a_1} (\mathbf{z_0^{gt}}),\quad && \mathbf{z_1} =  g_{a_1} (\mathbf{z_0}) && + d_1 (g_{a_1} (\mathbf{z_0})).
\end{alignedat}
\end{equation}
Note that the weights are shared for the parametric analysis $g_{a_0}$ and $g_{a_1}$ in two branches, and the denoisers are implemented in a residual manner. To enable direct supervision for feature denoising, a multiscale guidance loss is posed on the latent space to guide the learning of the denoisers. Specifically, the two-level guidance loss $\mathcal{G}$ is to minimize the $\mathcal{L}_1$ distance between the denoised and guiding features:
\begin{equation}
    \mathcal{G} = || \mathbf{z_0} - \mathbf{z_0^{gt}} ||_1 + || \mathbf{z_1} - \mathbf{z_1^{gt}} ||_1.
\end{equation}

\subsection{Rate-Distortion Optimization}
\label{rd-optimization}
Some entropy coding methods like arithmetic coding~\cite{rissanen1981universal} or asymmetric numeral systems (ANS)~\cite{duda2009asymmetric} are utilized to losslessly compress the discrete latent features $\mathbf{\hat{z}_1}$ and $\mathbf{\hat{z}_2}$ into bitstreams, which are the two parts of information needed to be stored during compression. As an inherent compression task, we hope the storing bitstreams are as short as possible; as a joint task with denoising, we hope to minimize the difference between the decoded image $\mathbf{\hat{x}}$ and the clean image $\mathbf{x}$. Hence, it is natural to apply the rate-distortion (RD) objective function $\mathcal{L}_{rd}$ in this joint task:
\begin{equation}
    \mathcal{L}_{rd} = \mathcal{R} (\mathbf{\hat{z}_1}) + \mathcal{R} (\mathbf{\hat{z}_2}) + \lambda_d \mathcal{D}(\mathbf{x}, \mathbf{\hat{x}}).
\end{equation}
In the context of rate-distortion theory, the to-be-coded sources are the noisy images $\mathbf{\tilde{x}}$, and the distortion is measured with respect to the corresponding clean counterparties $\mathbf{x}$. Similar to~\cite{cheng2020learned,minnen2018joint}, the rate $\mathcal{R}$ denotes the rate levels for the bitstreams, which is defined as the entropy of the latent variables:
\begin{equation}
\begin{split}
    & \mathcal{R} (\mathbf{\hat{z}_1}) = \mathbb{E}_{\mathbf{\tilde{x}} \sim p_\mathbf{\tilde{x}}}[-\log_2 p_{\mathbf{\hat{z}_1}|\mathbf{\hat{z}_2}}(\mathbf{\hat{z}_1}|\mathbf{\hat{z}_2})], \\
    & \mathcal{R} (\mathbf{\hat{z}_2}) = \mathbb{E}_{\mathbf{\tilde{x}} \sim p_\mathbf{\tilde{x}}}[-\log_2 p_{\mathbf{\hat{z}_2}|\theta}(\mathbf{\hat{z}_2}|\theta)].
\end{split}
\end{equation}
The formulation of the distortion $\mathcal{D}$ is different for MSE and MS-SSIM~\cite{wang2003msssim} optimizations, which is either $\mathcal{D} = \text{MSE}(\mathbf{x}, \mathbf{\hat{x}})$ or $\mathcal{D} = 1 − \text{MS-SSIM}(\mathbf{x}, \mathbf{\hat{x}})$. The factor $\lambda_d$ governs the trade-off between the bit rates $\mathcal{R}$ and the distortion $\mathcal{D}$.  

\subsection{Training Strategy}

\textbf{Pre-training as image compression.} 
Given that our joint image denoising and compression method is an inherent image compression algorithm, we first pre-train our network with only the guidance branch, where to-be-coded sources are the clean images $\mathbf{x}$ and the distortion is also measured with respect to $\mathbf{x}$. In this way, the compression capacity is enabled for our model with properly trained parameter weights, except for the denoiser $d$. In the supplements, we further present some ablation studies showing that the pre-training process on image compression can benefit and significantly boost the performance of the joint network.

\textbf{Fine-tuning under multiscale supervision.} 
The next step is to properly train the plug-in denoisers $d_1$, $d_2$ to enable the denoising capacity in the denoising branch. Specifically, noisy-clean image pairs are fed into the denoising and guidance branches accordingly for model fine-tuning, with both the rate-distortion loss $\mathcal{L}_{rd}$ and the guidance loss $\mathcal{G}$. In this way, the full objective function $\mathcal{L}$ during fine-tuning becomes
\begin{equation}
    \mathcal{L} = \mathcal{R} (\mathbf{\hat{z}_1}) + \mathcal{R} (\mathbf{\hat{z}_2}) + \lambda_d \mathcal{D}(\mathbf{x}, \mathbf{\hat{x}}) + \lambda_g \mathcal{G} (\mathbf{z_0}, \mathbf{z_0^{gt}}, \mathbf{z_1}, \mathbf{z_1^{gt}}),
\end{equation}
where $\lambda_g = 3.0$ is empirically set as the weight factor for the guidance loss.

\section{Experiments}

\subsection{Experimental Setup}
\textbf{Synthetic datasets.}
The Flicker 2W dataset~\cite{liu2020unified} is used for training and validation, which consists of $20,745$ general clean images. Similar to~\cite{xie2021enhanced}, images smaller than $256$ pixels are dropped for convenience, and around $200$ images are selected for validation. The Kodak PhotoCD image dataset (Kodak)~\cite{kodak} and the CLIC Professional Validation dataset (CLIC)~\cite{clic} are used for testing, which are two common datasets for the image compression task. There are $24$ high-quality $768 \times 512$ images in the Kodak dataset and $41$ higher-resolution images in the CLIC dataset.

We use the same noise sampling strategy as in~\cite{mildenhall2018kpn} during training, where the readout noise parameter $\sigma_r$ and the shot noise parameter $\sigma_s$ are uniformly sampled from $[10^{−3}, 10^{−1.5}]$ and $[10^{−4}, 10^{−2}]$, respectively. As for the validation and testing, the $4$ pre-determined parameter pairs $(\sigma_r, \sigma_s)$\footnote{Gain $\propto 1 = (10^{−2.1}, 10^{−2.6})$, Gain $\propto 2 = (10^{−1.8}, 10^{−2.3})$, Gain $\propto 4 = (10^{−1.4}, 10^{−1.9})$, Gain $\propto 8 = (10^{−1.1}, 10^{−1.5})$.} in~\cite{mildenhall2018kpn}'s official test set are used. Please note that Gain $\propto 4$ (slightly noisier) and Gain $\propto 8$ (significantly noisier) levels are unknown to the network during training. We test at full resolution on the Kodak and CLIC datasets with pre-determined levels of noise added.

\textbf{Real-world datasets.}
The public SIDD-Medium~\cite{abdelhamed2018sidd} dataset, containing $320$ noisy-clean sRGB image pairs for training, is adopted to further validate our method on real-world noisy images. The SIDD-Medium dataset contains $10$ different scenes with $160$ scene instances (different cameras, ISOs, shutter speeds, and illuminance), where $2$ image pairs are selected from each scene instance. Following the same settings in image denoising tasks, the models are validated on the $1280$ patches in the SIDD validation set and tested on the SIDD benchmark patches by submitting the results to the SIDD website. 

\textbf{Training details.} For implementation, we use the anchor model~\cite{cheng2020learned} as our network architecture (without $d_1$ and $d_2$) and choose the bottlenect of a single residual attention block~\cite{cheng2020learned} for the plug-in denoisers $d_1$ and $d_2$. During training, the network is optimized using randomly cropped patches at a resolution of $256$ pixels. All the models are fine-tuned on the pre-trained anchor models~\cite{cheng2020learned} provided by the popular CompressAI PyTorch library~\cite{begaint2020compressai} using a single RTX 2080 Ti GPU. Some ablation studies on the utilized modules and the training strategy can be found in our supplements.

The networks are optimized using the Adam~\cite{kingma2015adam} optimizer with a mini-batch size of $16$ for $600$ epochs. The initial learning rate is set as $10^{-4}$ and decayed by a factor of $0.1$ at epoch $450$ and $550$. Some typical techniques are utilized to avoid model collapse due to the random-initialized denoisers at the start of the fine-tuning process: 1) We warm up the fine-tuning process for the first $20$ epochs. 2) We have a loss cap for each model so that the network will skip the optimization of a mini step if the training loss is beyond the set threshold value.

We select the same hyperparameters as in~\cite{cheng2020learned} to train compression models target a high compression ratio (relatively low bit rate) for practical reasons. Lower-rate models ($q_1$, $q_2$, $q_3$) have channel number $N=128$, usually accompanied with smaller $\lambda_d$ values. The channel number $N$ is set as $192$ for higher-rate models ($q_4$, $q_5$, $q_6$) and optimized using larger $\lambda_d$ values. We train our MSE models under all the $6$ qualities, with $\lambda_d$ selected from the set $\{0.0018$, $0.0035$, $0.0067$, $0.0130$, $0.0250$, $0.0483\}$; the corresponding $\lambda_d$ values for MS-SSIM ($q_2$, $q_3$, $q_5$, $q_6$) are chosen from $\{4.58$, $8.73$, $31.73$, $60.50\}$.

\textbf{Evaluation metrics.} For the evaluation of rate-distortion (RD) performance, we use the peak signal-to-noise ratio (PSNR) and the multiscale structural similarity index (MS-SSIM)~\cite{wang2003msssim} with the corresponding bits per pixel (bpp). The RD curves are utilized to show the denoising and coding capacity of various models, where the MS-SSIM metric is converted to $−10\log_{10}(1−\text{MS-SSIM})$ as prior work~\cite{cheng2020learned} for better visualization. 

\begin{figure}[t]
        \includegraphics[width=0.48\linewidth]{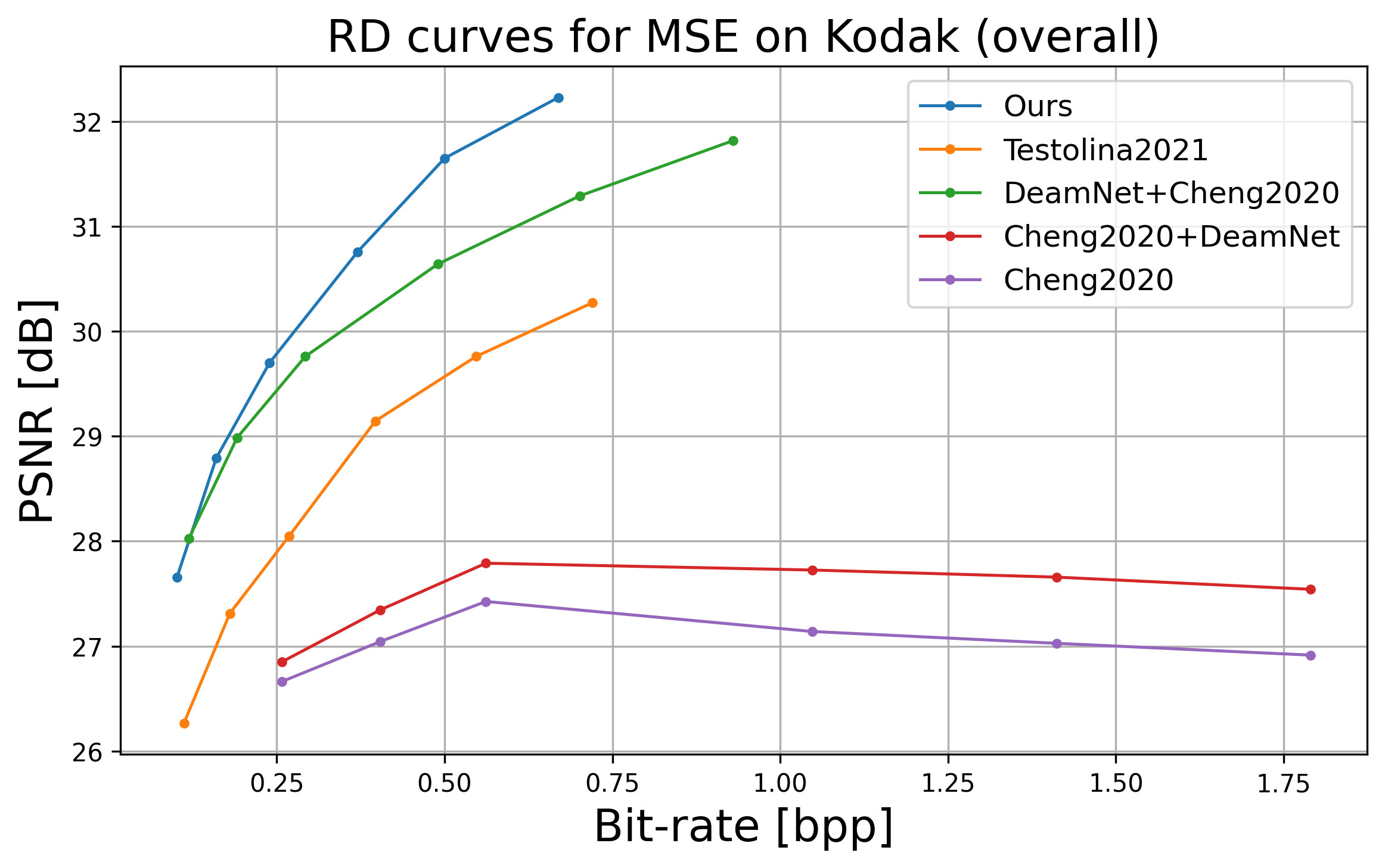}
        \includegraphics[width=0.48\linewidth]{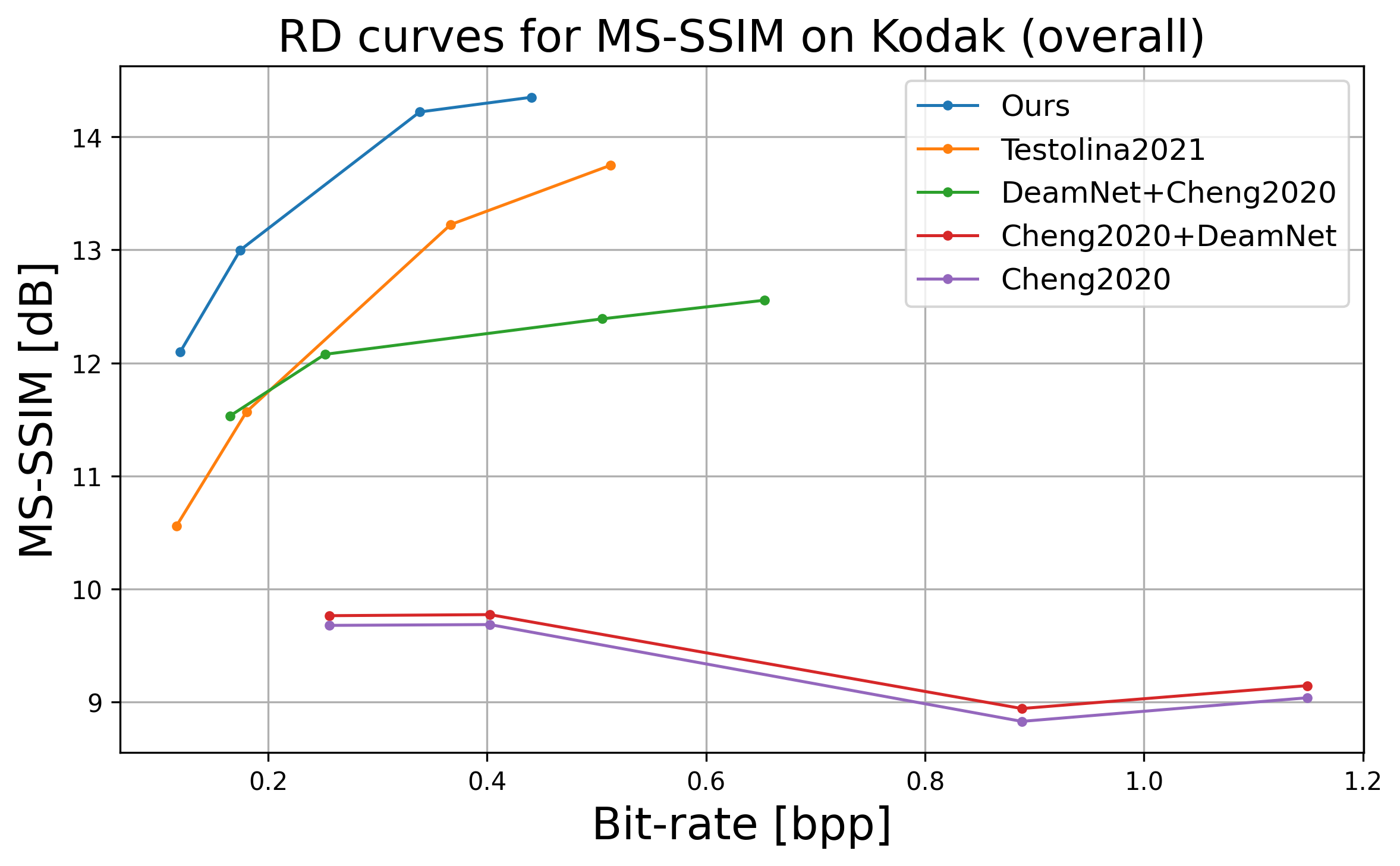}
\caption{Overall RD curves on the Kodak dataset at all noise levels. Our method has better RD performance over the pure compression, the sequential, and the joint baseline methods.}
\label{fig:rd_kodal_full}
\end{figure}

\begin{figure}[t]
        \includegraphics[width=0.48\linewidth]{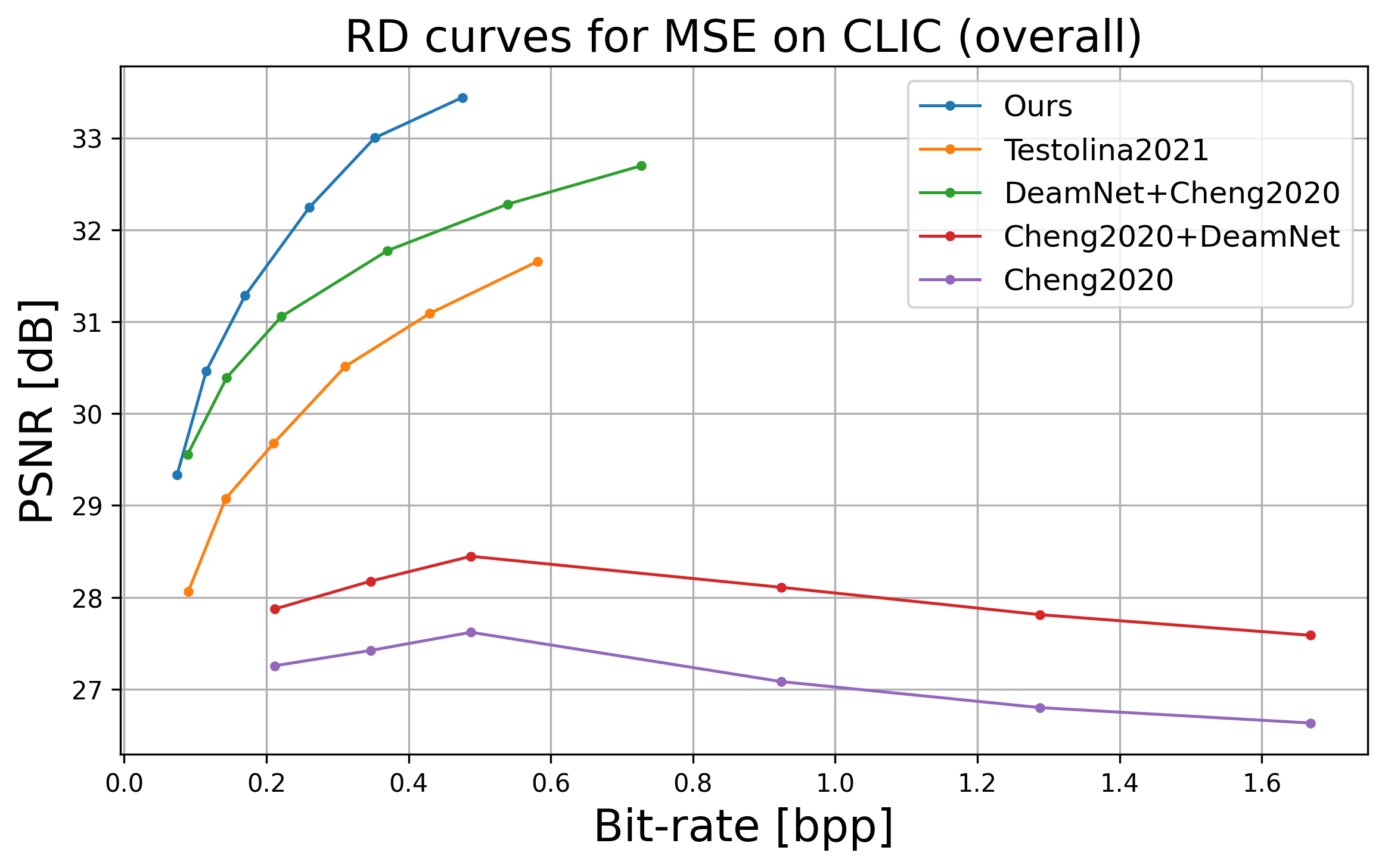}
        \includegraphics[width=0.48\linewidth]{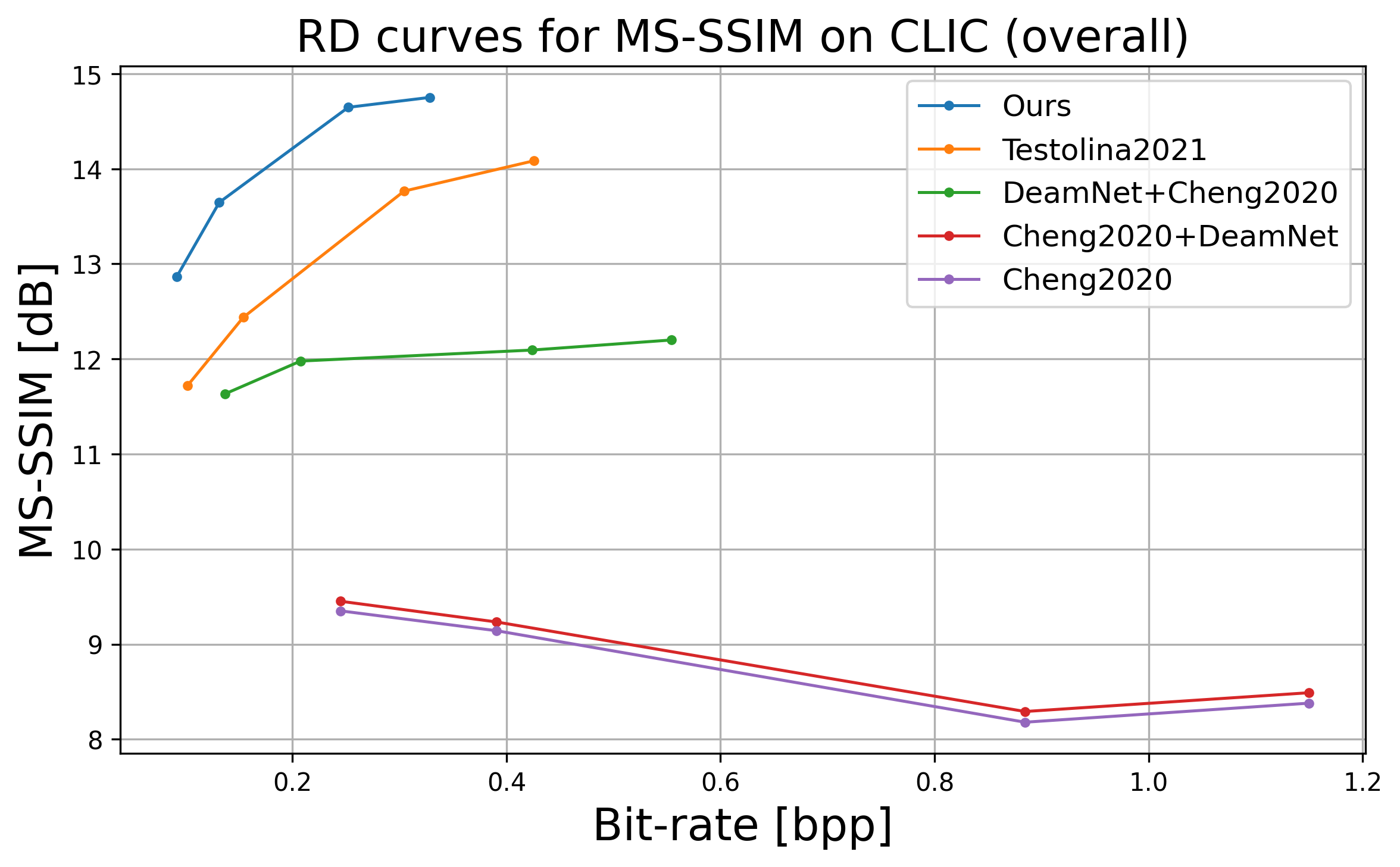}
\caption{Overall RD curves on the CLIC dataset at all noise levels. Our method has better RD performance over the pure compression, the sequential, and the joint baseline methods.}
\label{fig:rd_clic_full}
\end{figure}

\begin{figure}[t]
    \begin{subfigure}{1.0\linewidth}
        \includegraphics[width=0.48\linewidth]{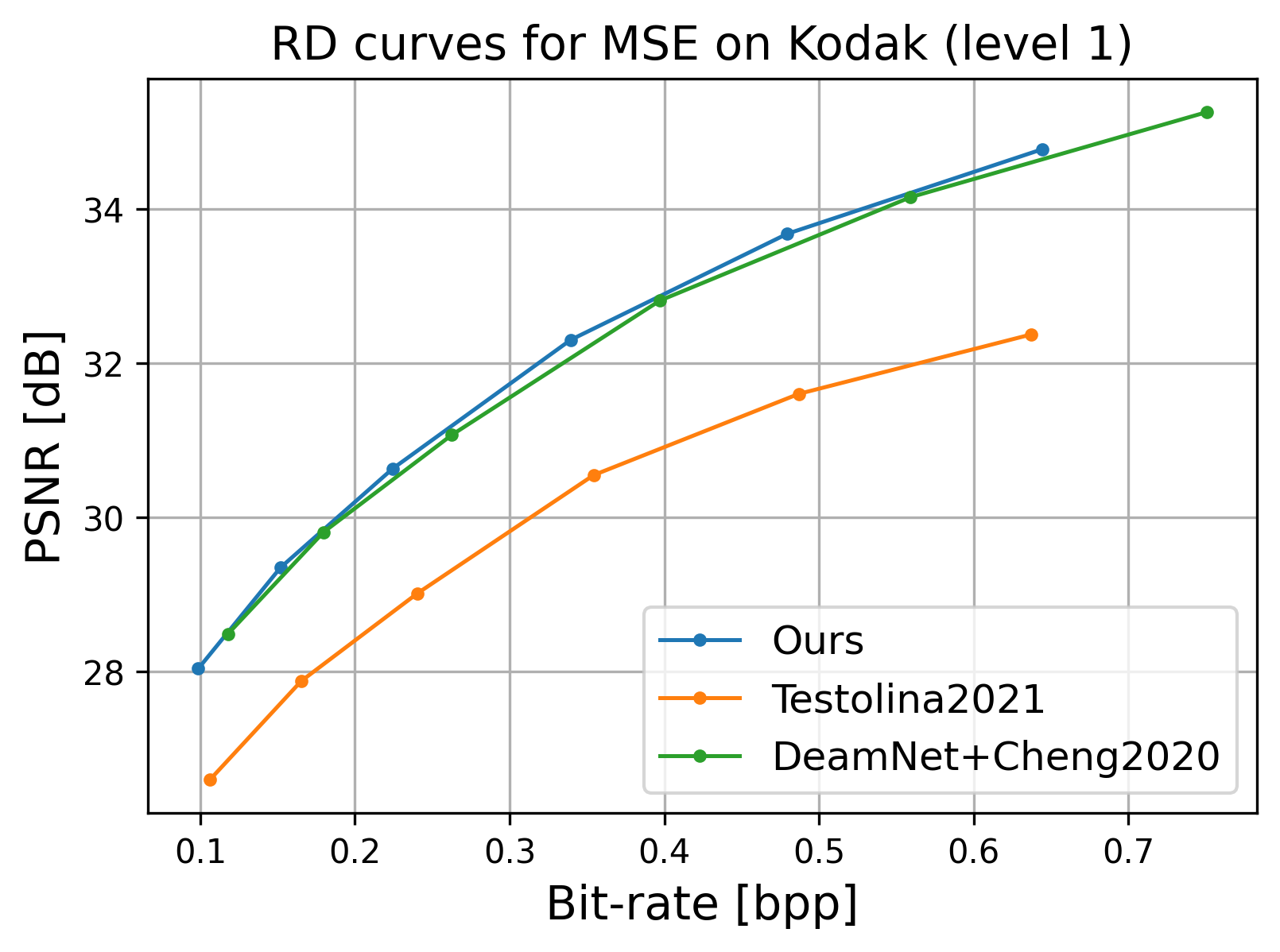}
        \includegraphics[width=0.48\linewidth]{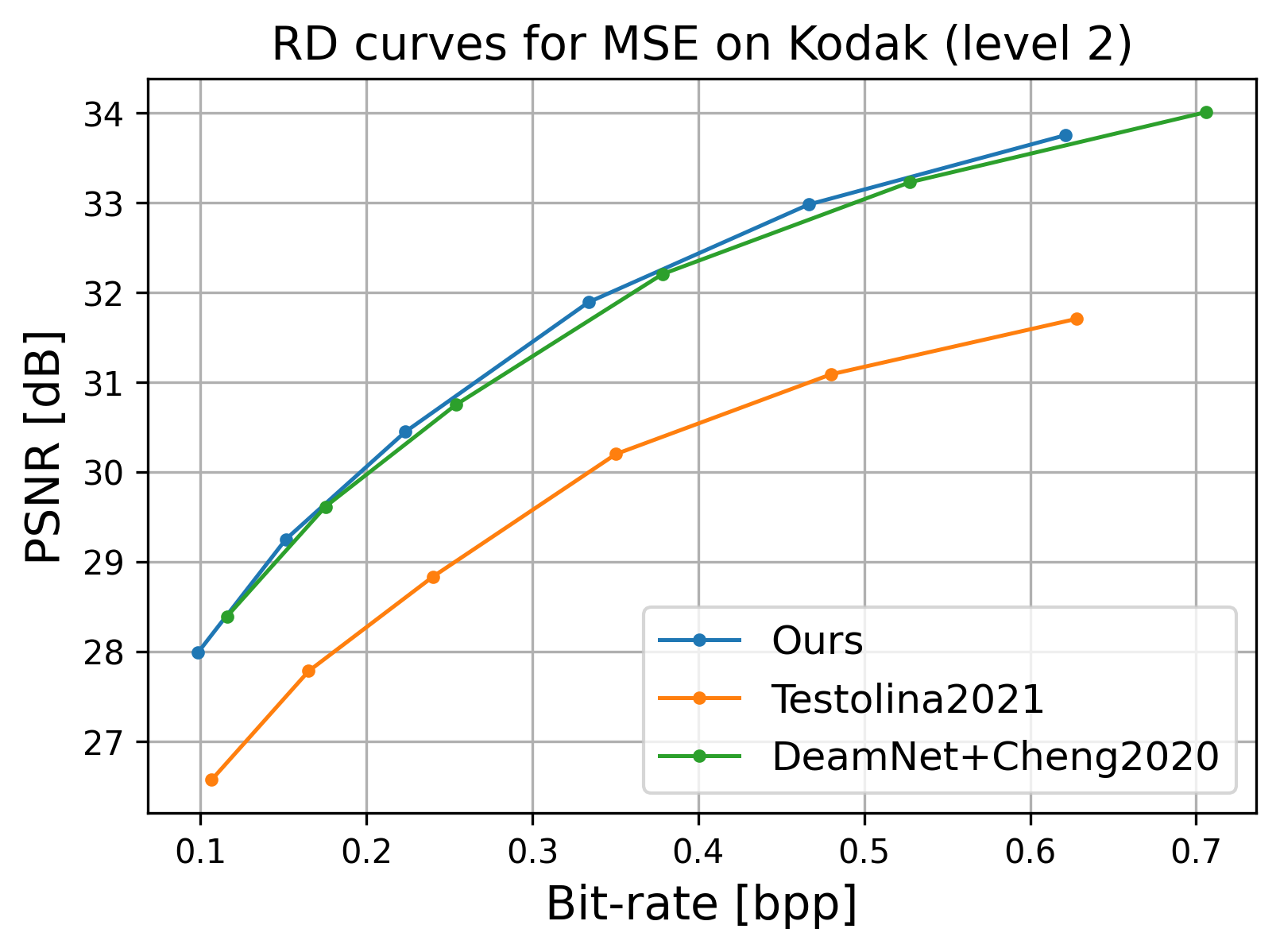}
    \end{subfigure}
    \begin{subfigure}{1.0\linewidth}
        \includegraphics[width=0.48\linewidth]{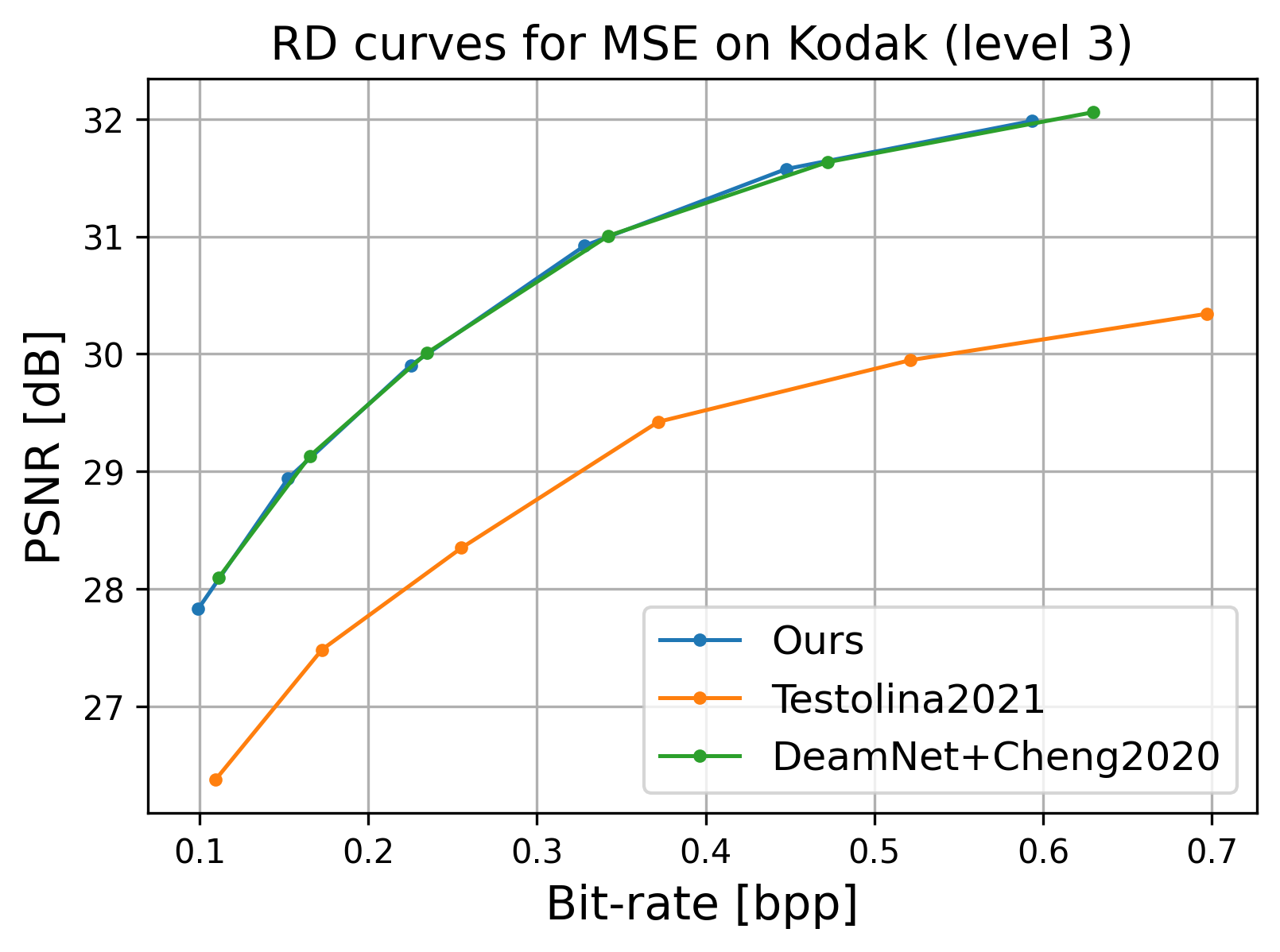}
        \includegraphics[width=0.48\linewidth]{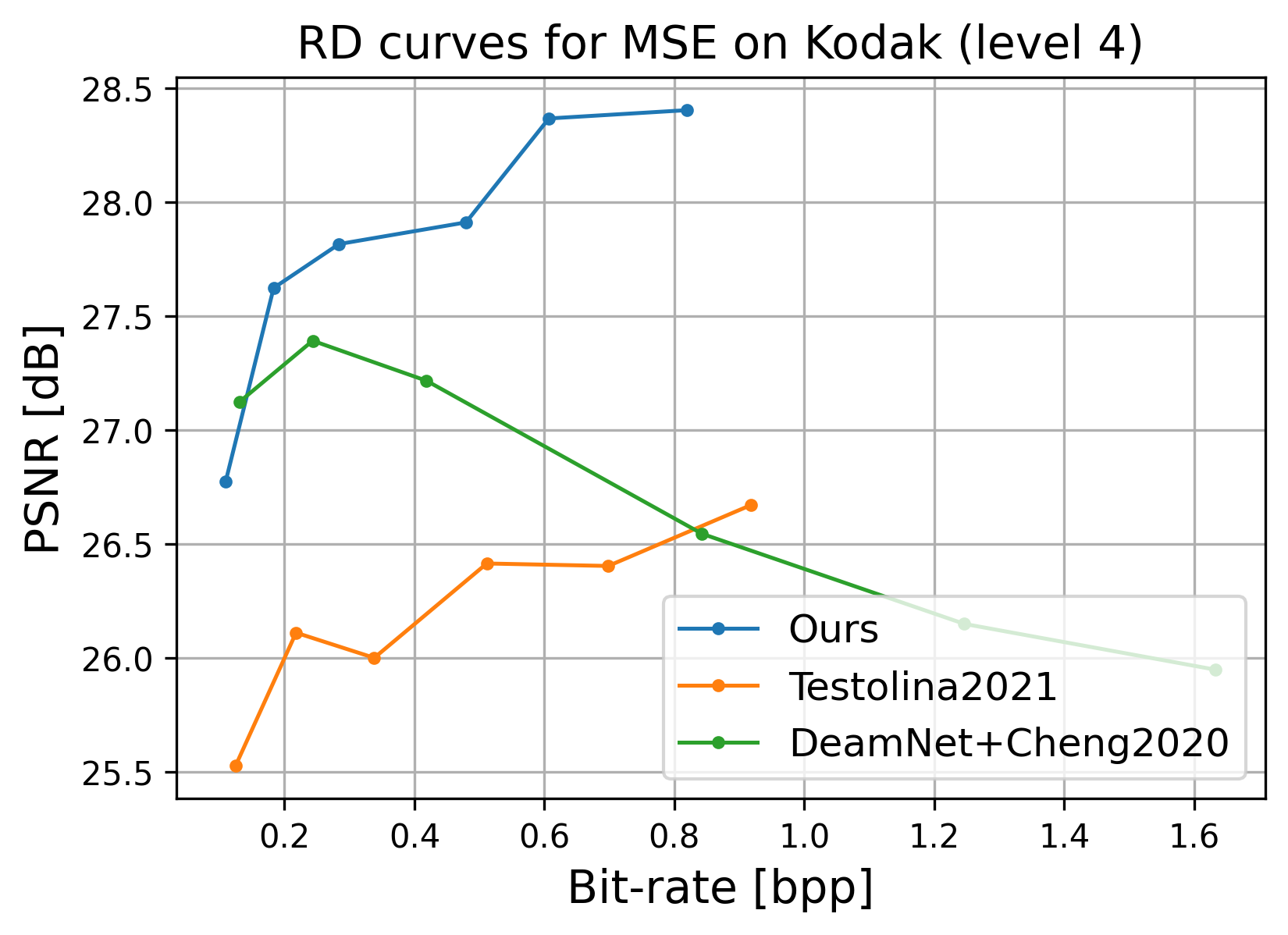}
    \end{subfigure}
\caption{RD curves on the Kodak dataset at individual noise level. Our method outperforms the baseline solutions, especially at the highest noise level.}
\label{fig:rd_kodak_partial}
\end{figure}

\begin{figure}[t]
    \begin{subfigure}{1.0\linewidth}
        \centering\includegraphics[width=0.6\linewidth]{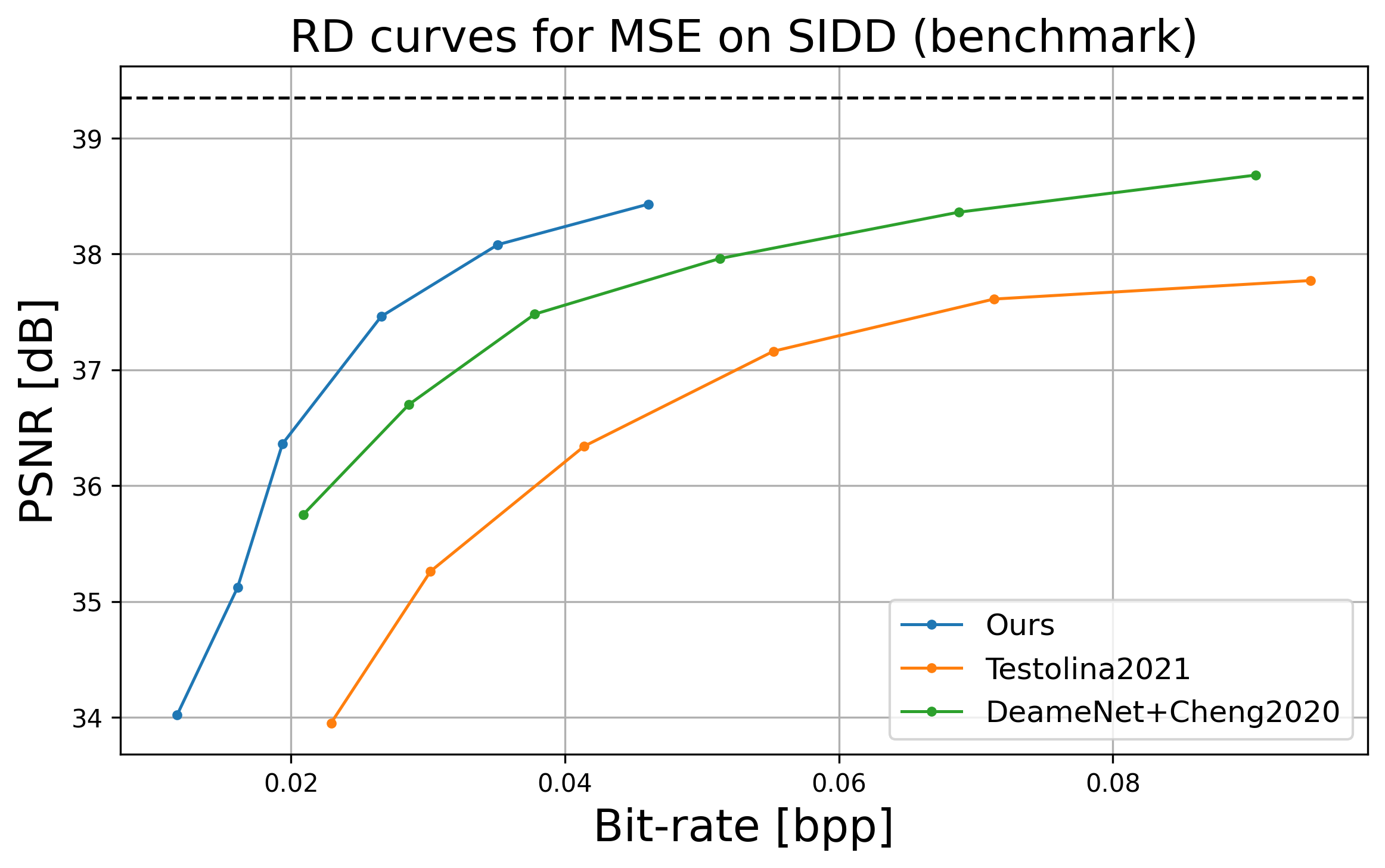}
    \end{subfigure}%
\caption{RD curves optimized for MSE on the SIDD. Our method outperforms all the baseline solutions. The black dotted line is the DeamNet ideal case without compression for reference.}
\label{fig:rd_sidd_partial}
\end{figure}

\subsection{Rate-Distortion Performance}
The sequential methods contains individual models of the state-of-the-art denoising DeamNet~\cite{ren2021adaptive} and the anchor compression model Cheng2020~\cite{cheng2020learned}. We compare our method with the following baseline methods: 1) ``Cheng2020+DeamNet'': sequential method of Cheng2020 and DeamNet; 2) ``DeamNet+Cheng2020'': sequential method of DeamNet and Cheng2020; 3) ``Testolina2021'': the joint baseline method~\cite{testolina2021towards}. We also report the performance of the pure image compression model ``Cheng2020'' on noisy-clean image pairs. Note that since a pure image compression model is trained to faithfully reconstruct the input image and is not expected to do any extra noisy-to-clean mapping, ``Cheng2020'' is only for qualitatively demonstrating the limitation of the current pure compression models as a reference.

For compression models, we use the pre-trained model provided by CompressAI~\cite{begaint2020compressai}. For the denoising models on SIDD, we use the officially pre-trained DeamNet; for models on synthetic data, we retrain DeamNet from stretch on the same synthetic training data as ours. We re-implement ``Testolina2021'' as the joint baseline method according to their original paper~\cite{testolina2021towards}. The RD results are obtained from the CompressAI evaluation platform and the official SIDD website. More quantitative results are available in our supplements.

\textbf{Synthetic noise (overall).} We show the overall (containing all the $4$ noise levels) RD curves for both the MSE and MS-SSIM methods evaluated on the Kodak dataset in Fig.~\ref{fig:rd_kodal_full} and on the CLIC dataset in Fig.~\ref{fig:rd_clic_full}. We can observe that our method (the blue RD curves) yields much better overall performance than the pure compression method, the sequential methods, and the joint baseline method.

For sequential methods, the green and red RD curves show that both sequential solutions have inferior performance compared to our joint solution. The execution order of the individual methods also matters. Intuitively, the sequential method that performs compression and successively denoising can suffer from the information loss and waste of bits allocating to image noise caused by the bottleneck of the existing general image compression method (see the purple RD curves for reference). The compressed noisy image with information loss makes the successive denoiser harder to reconstruct a pleasing image. Hence, in our remaining discussions, the sequential method specifically refers to the one that does denoising and successive compression. 

The orange RD curves show that the joint baseline method~\cite{testolina2021towards} cannot outperform the sequential one and have a more significant performance gap between our method due to the better design of our compressor to learn a noise-free representation compared to previous works.

\textbf{Synthetic noise (individual).} To further discuss the effects of different noise levels, Fig.~\ref{fig:rd_kodak_partial} shows the RD curves at individual noise levels for the MSE models on the Kodak dataset. We can see that our joint method is slightly better than the sequential method at the first three noise levels and significantly outperforms the sequential one at the highest noise level. Not to mention that our method has a much lower inference time as detailed in Sec.~\ref{speed}. 

It is interesting to know that the pure denoiser DeamNet (black dotted line) drops significantly down to around $24$ PSNR at noise level $4$, which is the direct cause of the degraded performance for the sequential method (green curve) in the fourth chart in Fig.~\ref{fig:rd_kodak_partial}. Recall that all the models are not trained on synthetic images at noise level $3$ (Gain $\propto 4$) and $4$ (Gain $\propto 8$), where the Gain $\propto 4$ noise is slightly higher while Gain $\propto 8$ noise is considerably higher than the noisiest level during training. This indicates that the performance of the sequential solutions is somehow limited by the capacity of individual modules and suffers from the accumulation of errors. Our joint method has a beneficial generalization property to the unseen noise level to a certain extent. 

\textbf{Real-world noise.} We also provide the RD curves optimized for MSE on the SIDD with real-world noise in Fig.~\ref{fig:rd_sidd_partial}. We plot DeamNet (black dotted line) as a pure denoising model to show an ideal case of denoising performance without compression (at $24$ bpp) for reference. The results show that our proposed method works well not only on the synthetic dataset but also on the images with real-world noise.

It is worth mentioning that given the same compressor, the compressed bit lengths of different images vary, depending on the amount of information (entropy) inside the images. Here, we can see that all the evaluated RD points are positioned in the very low bpp range ($< 0.1$ bpp). The very low bit-rate SIDD results are consistent among all methods, indicating inherently low entropy in the test samples, where the official SIDD test patches of size $256 \times 256$ contain relatively simple patterns. 

\begin{figure}[t!]
    \begin{subfigure}{0.800\linewidth}
        \begin{subfigure}{1.0\linewidth}
            \begin{subfigure}{0.2\linewidth}
    			\includegraphics[width=0.98\linewidth]{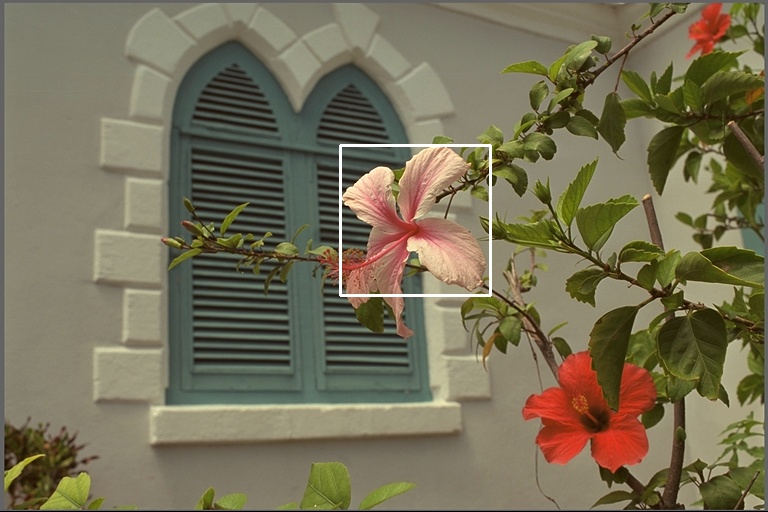}
        	\end{subfigure}%
            \begin{subfigure}{0.2\linewidth}
    			\includegraphics[width=0.98\linewidth]{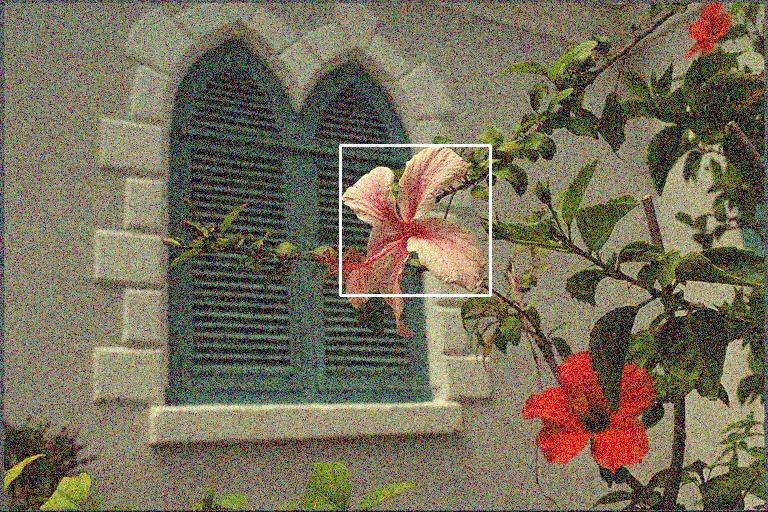}
        	\end{subfigure}%
            \begin{subfigure}{0.2\linewidth}
    			\includegraphics[width=0.98\linewidth]{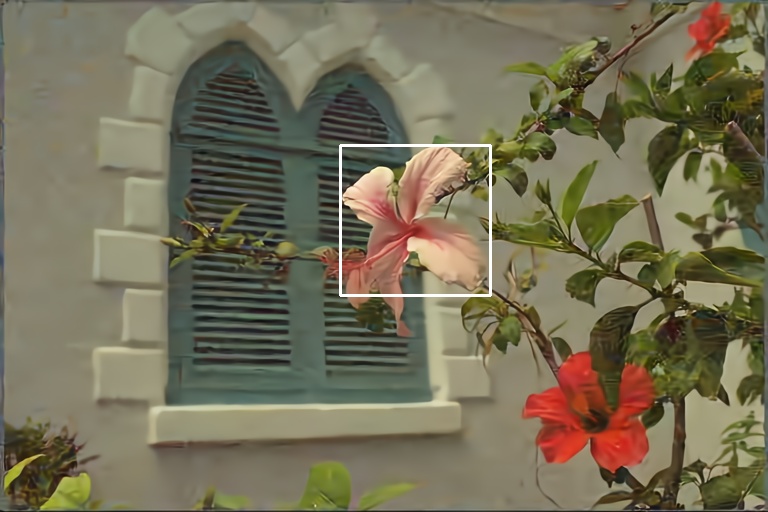}
        	\end{subfigure}%
            \begin{subfigure}{0.2\linewidth}
    			\includegraphics[width=0.98\linewidth]{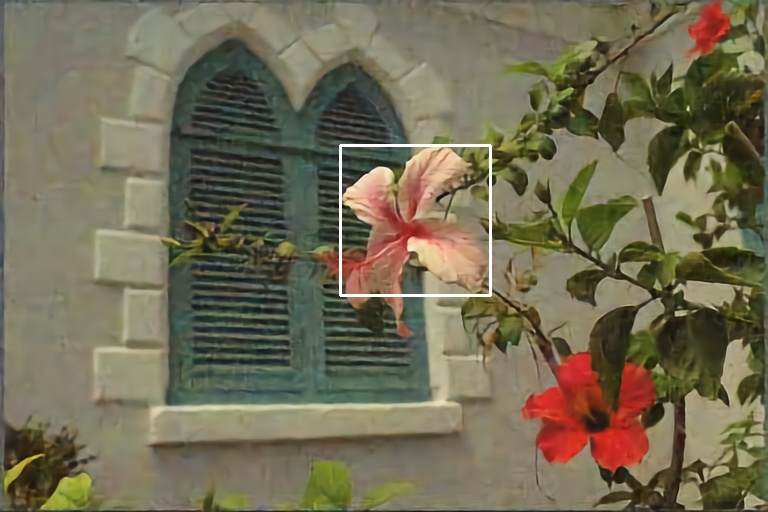}
        	\end{subfigure}%
            \begin{subfigure}{0.2\linewidth}
    			\includegraphics[width=0.98\linewidth]{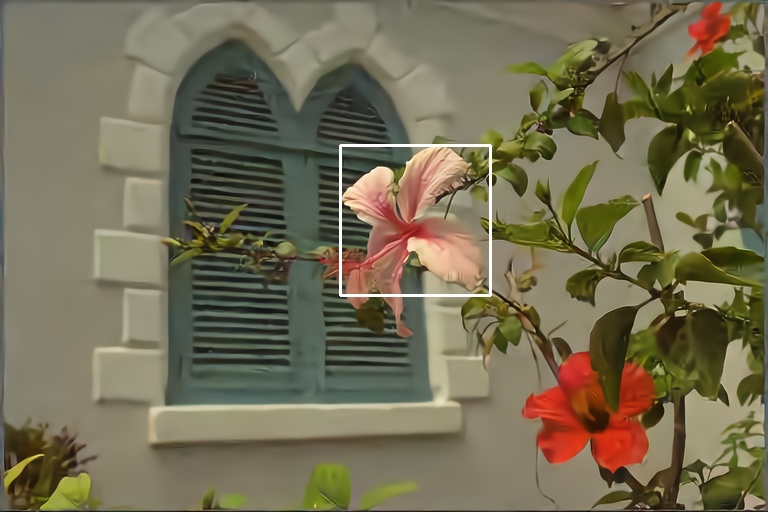}
        	\end{subfigure}%
    	\end{subfigure}
    	\\[2pt]
        \begin{subfigure}{1.0\linewidth}
            \begin{subfigure}{0.19\linewidth}\centering\scriptsize GT\end{subfigure}
            \begin{subfigure}{0.19\linewidth}\centering\scriptsize Noisy\end{subfigure}
            \begin{subfigure}{0.19\linewidth}\centering\scriptsize Sequential\\(28.816dB, 0.1859bpp)\end{subfigure}
            \begin{subfigure}{0.19\linewidth}\centering\scriptsize Baseline\\(27.402dB, 0.2169bpp)\end{subfigure}
            \begin{subfigure}{0.19\linewidth}\centering\scriptsize Ours\\(28.916dB, 0.1502bpp)\end{subfigure}
        \end{subfigure}
    \end{subfigure}
    \begin{subfigure}{0.19\linewidth}
        \begin{subfigure}{1.0\linewidth}
            \begin{subfigure}{0.08\linewidth}
                \hfil\parbox[][][c]{\linewidth}{\centering\scriptsize\raisebox{0in}{\rotatebox{90}{GT}}}
            \end{subfigure}
            \begin{subfigure}{0.32\linewidth}
                \includegraphics[width=0.98\linewidth]{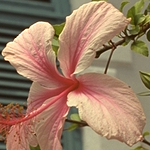}
            \end{subfigure}
            \begin{subfigure}{0.08\linewidth}
                \hfil\parbox[][][c]{\linewidth}{\centering\scriptsize\raisebox{0in}{\rotatebox{90}{Noisy}}}
            \end{subfigure}
            \begin{subfigure}{0.32\linewidth}
                \includegraphics[width=0.98\linewidth]{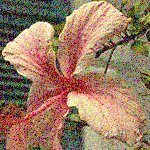}
            \end{subfigure}
        \end{subfigure}
        \\[2pt]
        \begin{subfigure}{1.0\linewidth}
            \begin{subfigure}{0.31\linewidth}
               \includegraphics[width=0.98\linewidth]{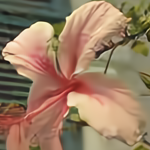}
            \end{subfigure}
            \begin{subfigure}{0.31\linewidth}
                \includegraphics[width=0.98\linewidth]{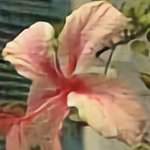}
            \end{subfigure}
            \begin{subfigure}{0.31\linewidth}
                \includegraphics[width=0.98\linewidth]{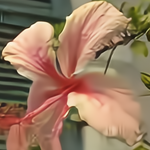}
            \end{subfigure}
        \end{subfigure}
        \\[2pt]
        \begin{subfigure}{1.0\linewidth}
            \begin{subfigure}{0.31\linewidth}\centering\scriptsize Seq.\end{subfigure}
            \begin{subfigure}{0.31\linewidth}\centering\scriptsize Base.\end{subfigure}
            \begin{subfigure}{0.31\linewidth}\centering\scriptsize Ours\end{subfigure}
        \end{subfigure}
    \end{subfigure}
    \\[2pt]
    \begin{subfigure}{0.800\linewidth}
        \begin{subfigure}{1.0\linewidth}
            \begin{subfigure}{0.2\linewidth}
    			\includegraphics[width=0.98\linewidth]{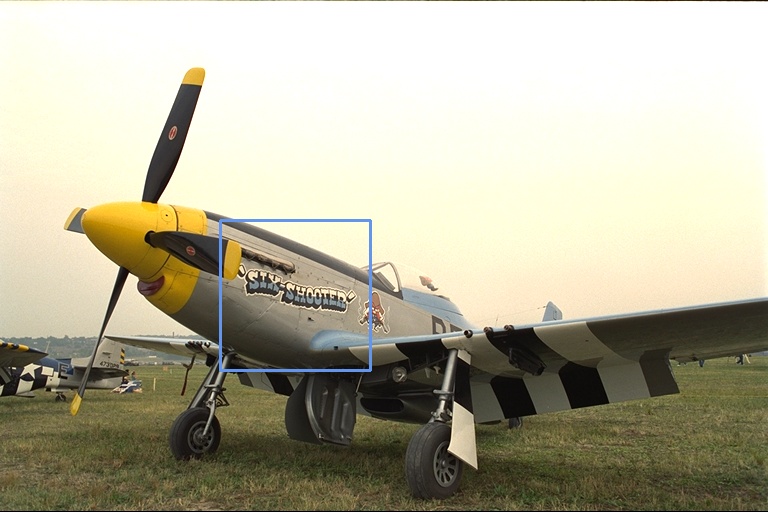}
        	\end{subfigure}%
            \begin{subfigure}{0.2\linewidth}
    			\includegraphics[width=0.98\linewidth]{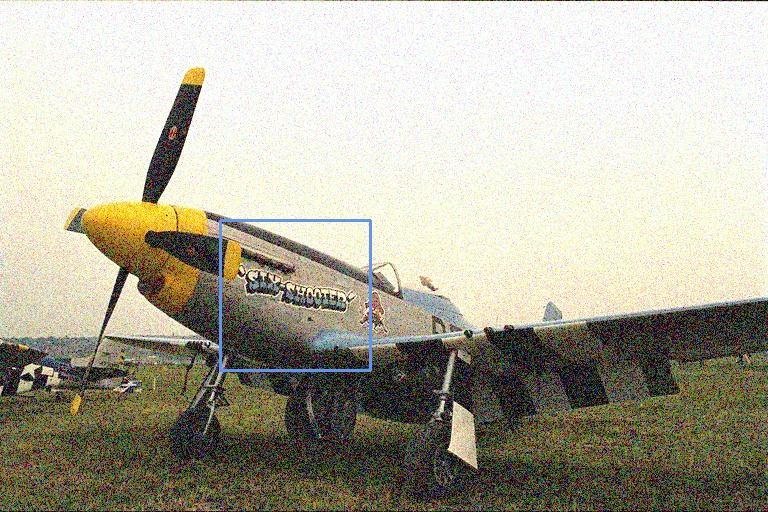}
        	\end{subfigure}%
            \begin{subfigure}{0.2\linewidth}
    			\includegraphics[width=0.98\linewidth]{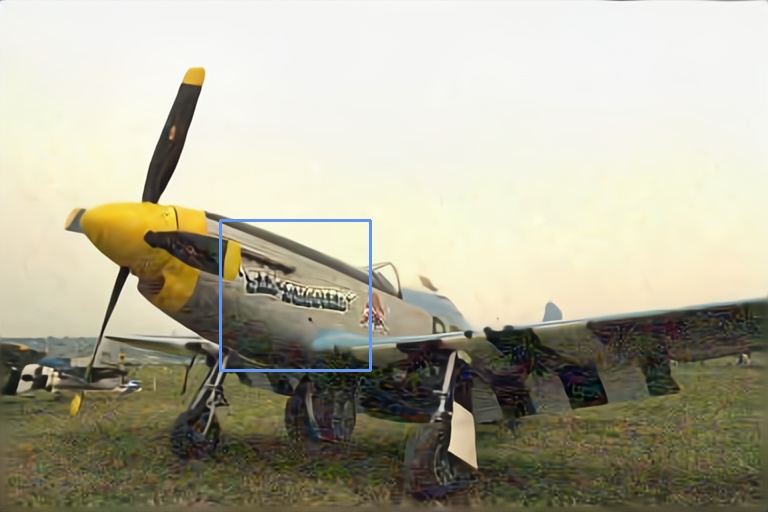}
        	\end{subfigure}%
            \begin{subfigure}{0.2\linewidth}
    			\includegraphics[width=0.98\linewidth]{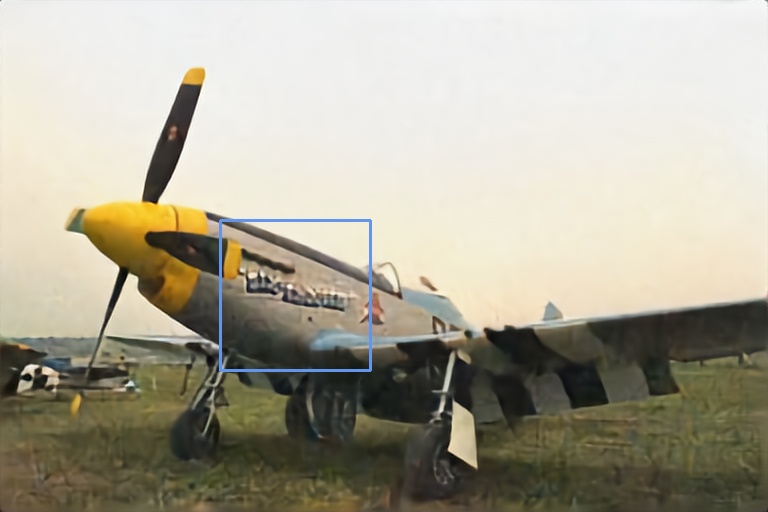}
        	\end{subfigure}%
            \begin{subfigure}{0.2\linewidth}
    			\includegraphics[width=0.98\linewidth]{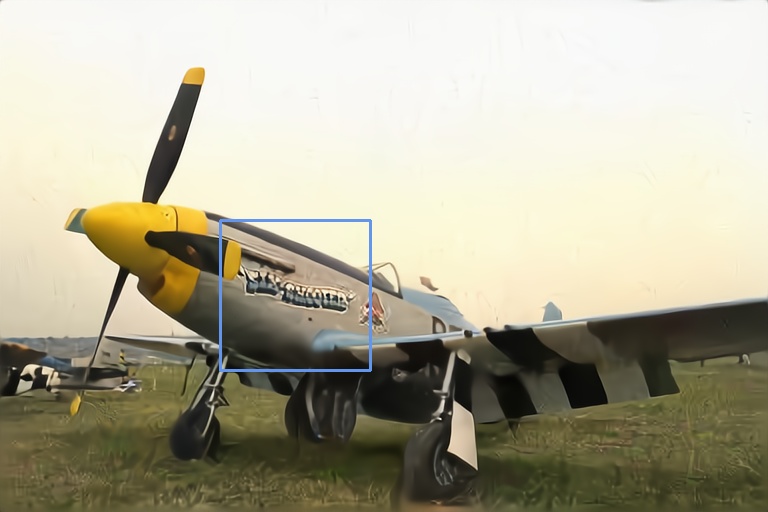}
        	\end{subfigure}%
    	\end{subfigure}
    	\\[2pt]
        \begin{subfigure}{1.0\linewidth}
            \begin{subfigure}{0.19\linewidth}\centering\scriptsize GT\end{subfigure}
            \begin{subfigure}{0.19\linewidth}\centering\scriptsize Noisy\end{subfigure}
            \begin{subfigure}{0.19\linewidth}\centering\scriptsize Sequential\\(0.9269, 0.1436bpp)\end{subfigure}
            \begin{subfigure}{0.19\linewidth}\centering\scriptsize Baseline\\(0.9382, 0.1366bpp)\end{subfigure}
            \begin{subfigure}{0.19\linewidth}\centering\scriptsize Ours\\(0.9503, 0.1045bpp)\end{subfigure}
        \end{subfigure}
    \end{subfigure}
    \begin{subfigure}{0.19\linewidth}
        \begin{subfigure}{1.0\linewidth}
            \begin{subfigure}{0.08\linewidth}
                \hfil\parbox[][][c]{\linewidth}{\centering\scriptsize\raisebox{0in}{\rotatebox{90}{GT}}}
            \end{subfigure}
            \begin{subfigure}{0.32\linewidth}
                \includegraphics[width=0.98\linewidth]{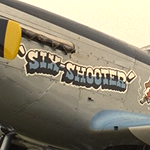}
            \end{subfigure}
            \begin{subfigure}{0.08\linewidth}
                \hfil\parbox[][][c]{\linewidth}{\centering\scriptsize\raisebox{0in}{\rotatebox{90}{Noisy}}}
            \end{subfigure}
            \begin{subfigure}{0.32\linewidth}
                \includegraphics[width=0.98\linewidth]{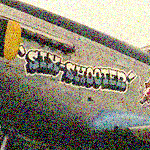}
            \end{subfigure}
        \end{subfigure}
        \\[2pt]
        \begin{subfigure}{1.0\linewidth}
            \begin{subfigure}{0.31\linewidth}
               \includegraphics[width=0.98\linewidth]{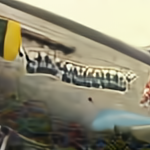}
            \end{subfigure}
            \begin{subfigure}{0.31\linewidth}
                \includegraphics[width=0.98\linewidth]{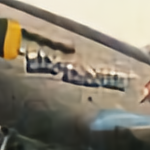}
            \end{subfigure}
            \begin{subfigure}{0.31\linewidth}
                \includegraphics[width=0.98\linewidth]{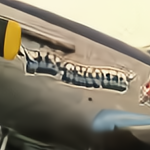}
            \end{subfigure}
        \end{subfigure}
        \\[2pt]
        \begin{subfigure}{1.0\linewidth}
            \begin{subfigure}{0.31\linewidth}\centering\scriptsize Seq.\end{subfigure}
            \begin{subfigure}{0.31\linewidth}\centering\scriptsize Base.\end{subfigure}
            \begin{subfigure}{0.31\linewidth}\centering\scriptsize Ours\end{subfigure}
        \end{subfigure}
    \end{subfigure}
    
	\caption{Comparison results at noise level $4$ (Gain $\propto 8$) on Kodak image $\textit{kodim07}$ for MSE models and on Kodak image $\textit{kodim20}$ for MS-SSIM models. Apart from the better PSNR values and lower bpp rates, we can see that our solution has a better capacity to restore structural texture and edges.}
	\label{fig:qualitatives_kodak}
\end{figure}

\begin{figure}[t!]
    \begin{subfigure}{0.800\linewidth}
        \begin{subfigure}{1.0\linewidth}
            \begin{subfigure}{0.2\linewidth}
    			\includegraphics[width=0.98\linewidth]{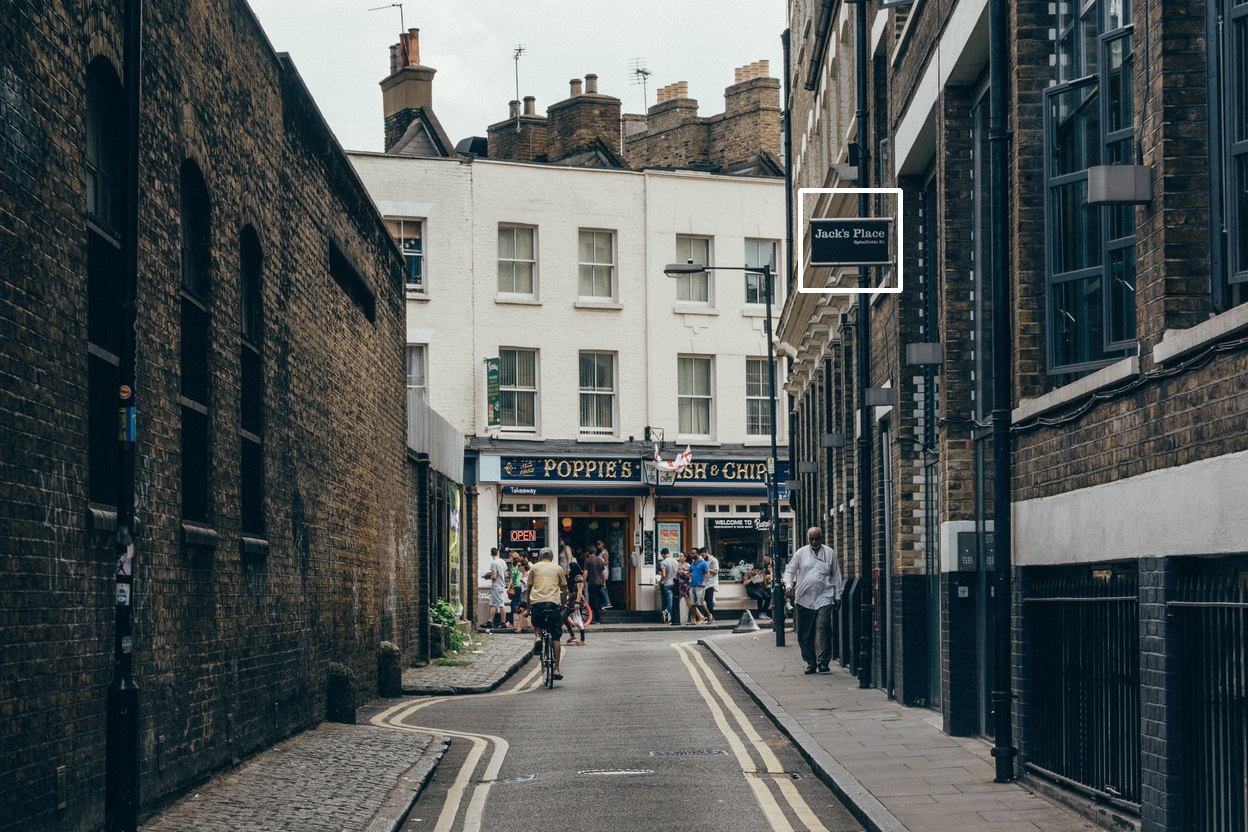}
        	\end{subfigure}%
            \begin{subfigure}{0.2\linewidth}
    			\includegraphics[width=0.98\linewidth]{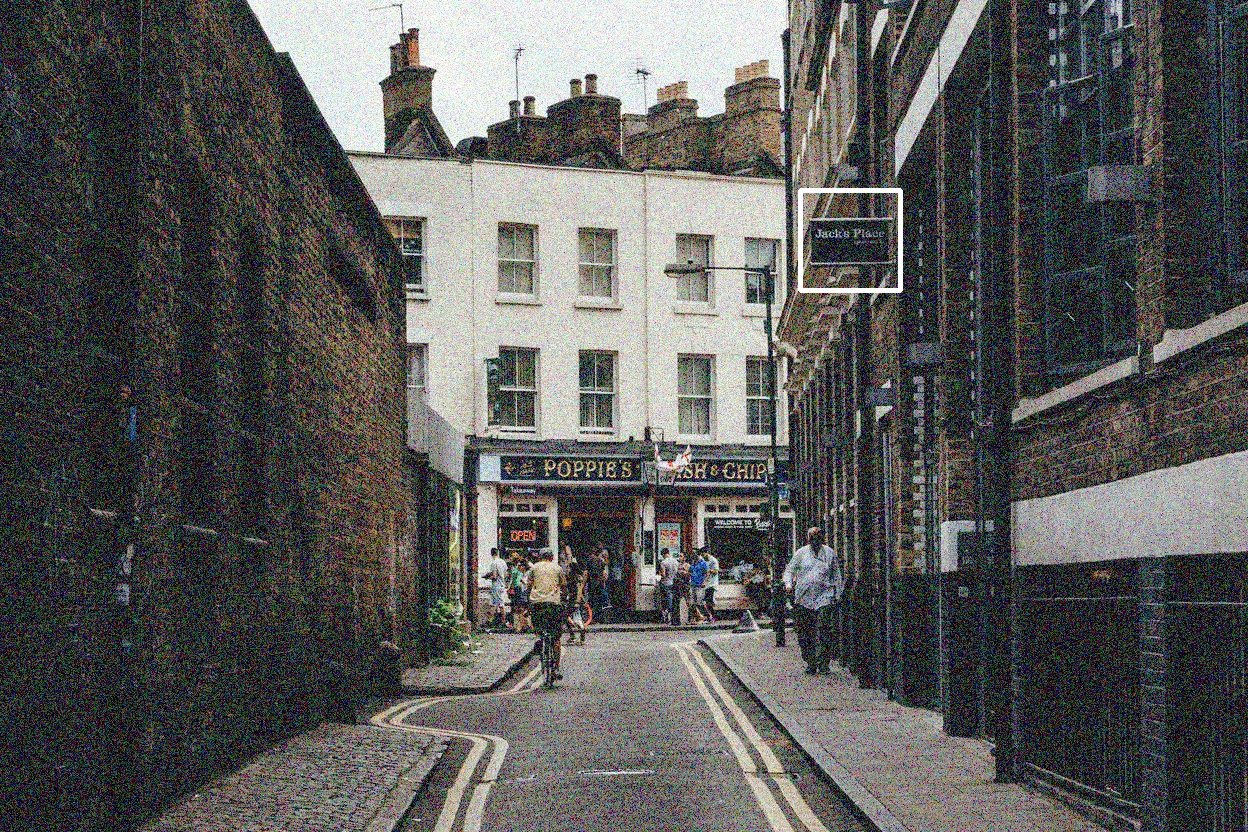}
        	\end{subfigure}%
            \begin{subfigure}{0.2\linewidth}
    			\includegraphics[width=0.98\linewidth]{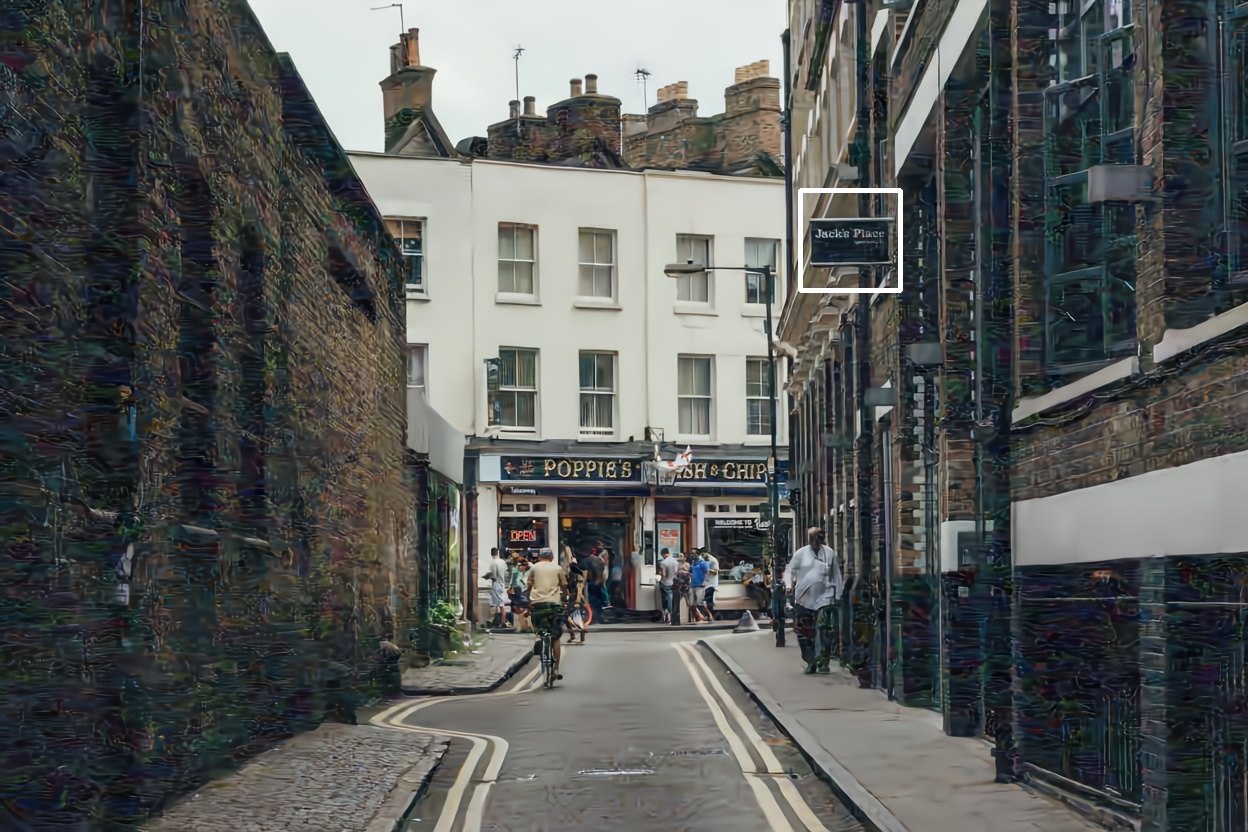}
        	\end{subfigure}%
            \begin{subfigure}{0.2\linewidth}
    			\includegraphics[width=0.98\linewidth]{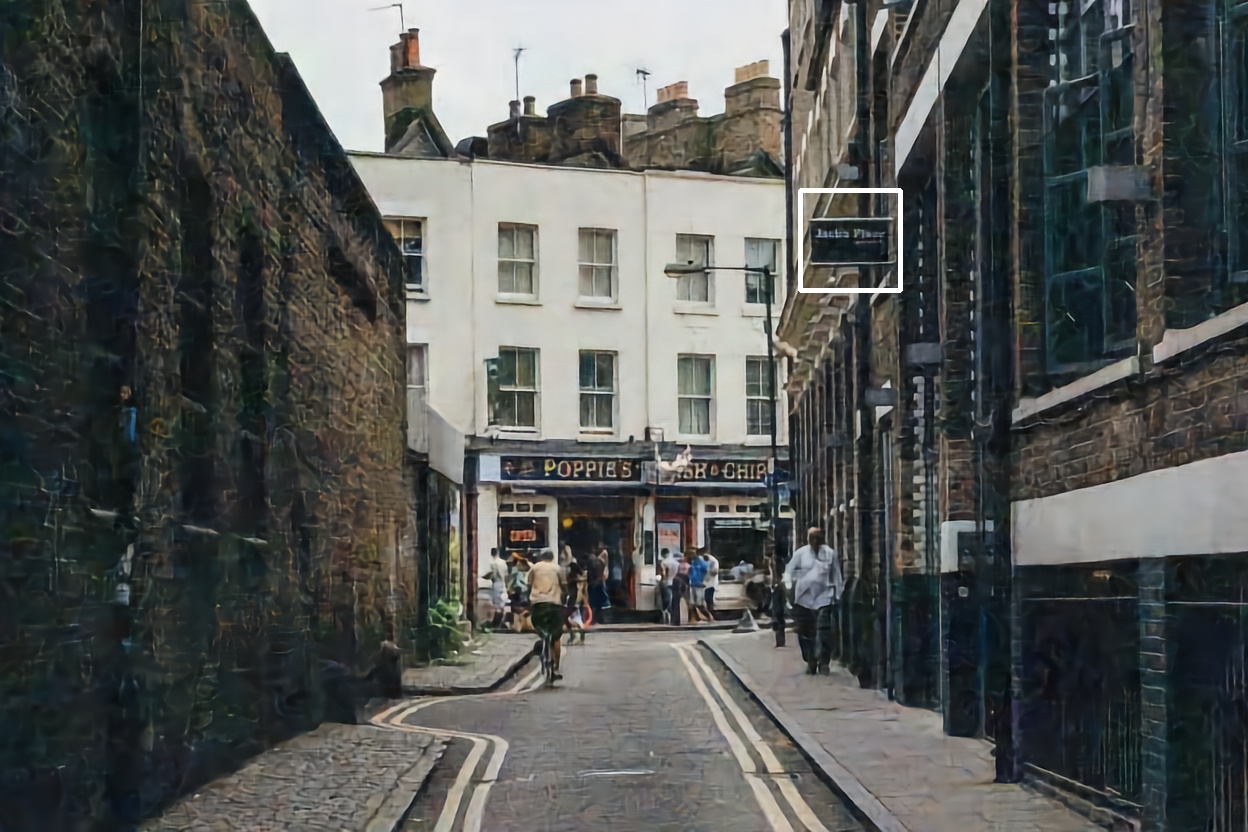}
        	\end{subfigure}
            \begin{subfigure}{0.2\linewidth}
    			\includegraphics[width=0.98\linewidth]{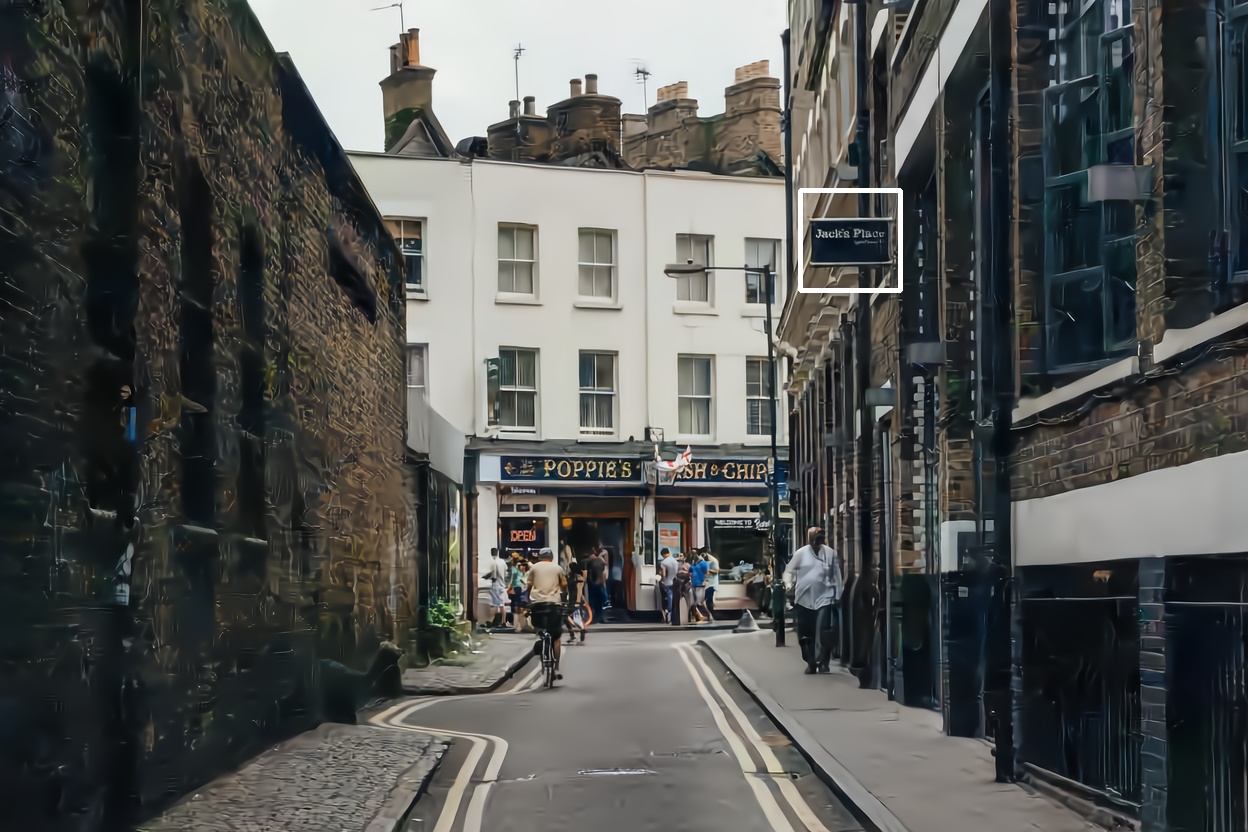}
        	\end{subfigure}%
    	\end{subfigure}
    	\\[2pt]
        \begin{subfigure}{1.0\linewidth}
            \begin{subfigure}{0.19\linewidth}\centering\scriptsize GT\end{subfigure}
            \begin{subfigure}{0.19\linewidth}\centering\scriptsize Noisy\end{subfigure}
            \begin{subfigure}{0.19\linewidth}\centering\scriptsize Sequential\\(25.005dB, 0.2908bpp)\end{subfigure}
            \begin{subfigure}{0.19\linewidth}\centering\scriptsize Baseline\\(25.249dB, 0.2230bpp)\end{subfigure}
            \begin{subfigure}{0.19\linewidth}\centering\scriptsize Ours\\(25.988dB, 0.1841bpp)\end{subfigure}
        \end{subfigure}
    \end{subfigure}
    \begin{subfigure}{0.19\linewidth}
        \begin{subfigure}{1.0\linewidth}
            \begin{subfigure}{0.01\linewidth} \hfil \end{subfigure}
            \begin{subfigure}{0.08\linewidth}
                \hfil\parbox[][][c]{\linewidth}{\centering\scriptsize\raisebox{0in}{\rotatebox{90}{GT}}}
            \end{subfigure}
            \begin{subfigure}{0.32\linewidth}
                \includegraphics[width=0.98\linewidth]{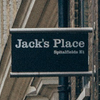}
            \end{subfigure}
            \begin{subfigure}{0.08\linewidth}
                \hfil\parbox[][][c]{\linewidth}{\centering\scriptsize\raisebox{0in}{\rotatebox{90}{Noisy}}}
            \end{subfigure}
            \begin{subfigure}{0.32\linewidth}
                \includegraphics[width=0.98\linewidth]{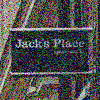}
            \end{subfigure}
        \end{subfigure}
        \\[2pt]
        \begin{subfigure}{1.0\linewidth}
            \begin{subfigure}{0.31\linewidth}
               \includegraphics[width=0.98\linewidth]{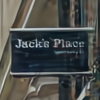}
            \end{subfigure}
            \begin{subfigure}{0.31\linewidth}
                \includegraphics[width=0.98\linewidth]{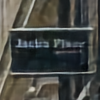}
            \end{subfigure}
            \begin{subfigure}{0.31\linewidth}
                \includegraphics[width=0.98\linewidth]{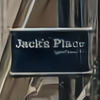}
            \end{subfigure}
        \end{subfigure}
        \\[2pt]
        \begin{subfigure}{1.0\linewidth}
            \begin{subfigure}{0.31\linewidth}\centering\scriptsize Seq.\end{subfigure}
            \begin{subfigure}{0.31\linewidth}\centering\scriptsize Base.\end{subfigure}
            \begin{subfigure}{0.31\linewidth}\centering\scriptsize Ours\end{subfigure}
        \end{subfigure}
    \end{subfigure}
    \\[2pt]
    \begin{subfigure}{0.800\linewidth}
        \begin{subfigure}{1.0\linewidth}
            \begin{subfigure}{0.2\linewidth}
    			\includegraphics[width=0.98\linewidth]{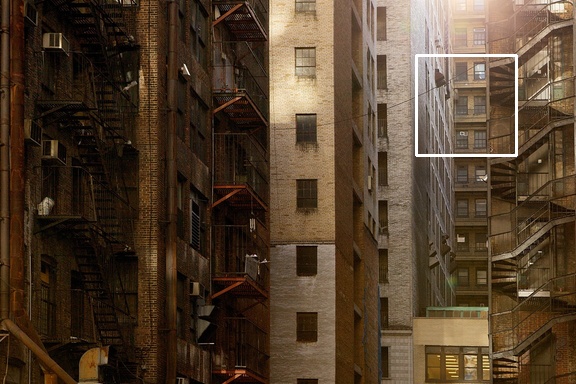}
        	\end{subfigure}%
            \begin{subfigure}{0.2\linewidth}
    			\includegraphics[width=0.98\linewidth]{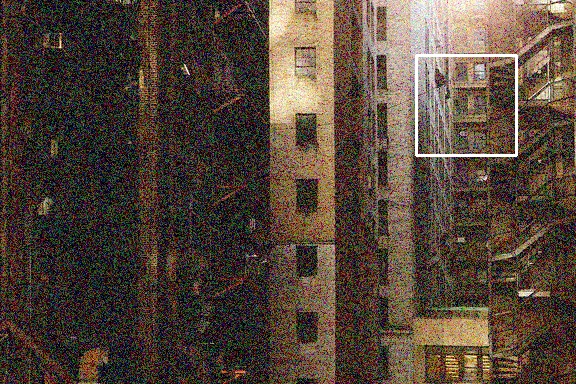}
        	\end{subfigure}%
            \begin{subfigure}{0.2\linewidth}
    			\includegraphics[width=0.98\linewidth]{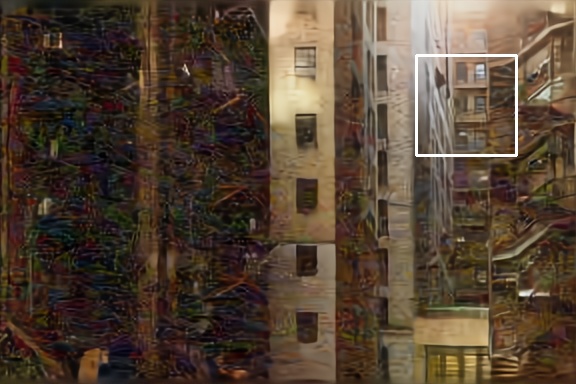}
        	\end{subfigure}%
            \begin{subfigure}{0.2\linewidth}
    			\includegraphics[width=0.98\linewidth]{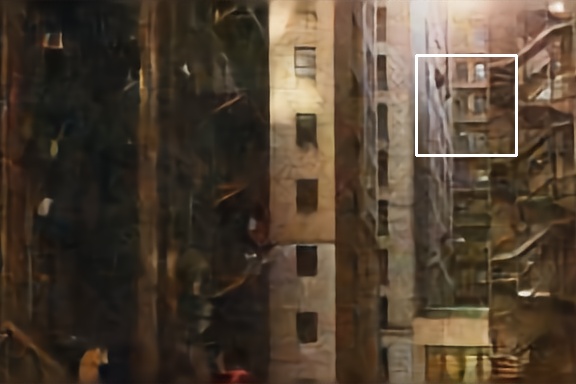}
        	\end{subfigure}
            \begin{subfigure}{0.2\linewidth}
    			\includegraphics[width=0.98\linewidth]{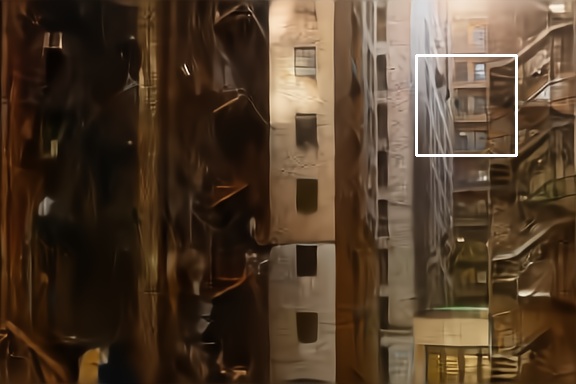}
        	\end{subfigure}%
    	\end{subfigure}
    	\\[2pt]
        \begin{subfigure}{1.0\linewidth}
            \begin{subfigure}{0.19\linewidth}\centering\scriptsize GT\end{subfigure}
            \begin{subfigure}{0.19\linewidth}\centering\scriptsize Noisy\end{subfigure}
            \begin{subfigure}{0.19\linewidth}\centering\scriptsize Sequential\\(0.8035, 0.2689bpp)\end{subfigure}
            \begin{subfigure}{0.19\linewidth}\centering\scriptsize Baseline\\(0.8483, 0.1912bpp)\end{subfigure}
            \begin{subfigure}{0.19\linewidth}\centering\scriptsize Ours\\(0.8832, 0.1618bpp)\end{subfigure}
        \end{subfigure}
    \end{subfigure}
    \begin{subfigure}{0.19\linewidth}
        \begin{subfigure}{1.0\linewidth}
            \begin{subfigure}{0.01\linewidth} \hfil \end{subfigure}
            \begin{subfigure}{0.08\linewidth}
                \hfil\parbox[][][c]{\linewidth}{\centering\scriptsize\raisebox{0in}{\rotatebox{90}{GT}}}
            \end{subfigure}
            \begin{subfigure}{0.32\linewidth}
                \includegraphics[width=0.98\linewidth]{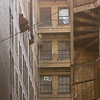}
            \end{subfigure}
            \begin{subfigure}{0.08\linewidth}
                \hfil\parbox[][][c]{\linewidth}{\centering\scriptsize\raisebox{0in}{\rotatebox{90}{Noisy}}}
            \end{subfigure}
            \begin{subfigure}{0.32\linewidth}
                \includegraphics[width=0.98\linewidth]{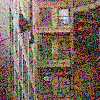}
            \end{subfigure}
        \end{subfigure}
        \\[2pt]
        \begin{subfigure}{1.0\linewidth}
            \begin{subfigure}{0.31\linewidth}
               \includegraphics[width=0.98\linewidth]{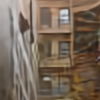}
            \end{subfigure}
            \begin{subfigure}{0.31\linewidth}
                \includegraphics[width=0.98\linewidth]{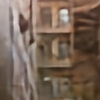}
            \end{subfigure}
            \begin{subfigure}{0.31\linewidth}
                \includegraphics[width=0.98\linewidth]{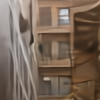}
            \end{subfigure}
        \end{subfigure}
        \\[2pt]
        \begin{subfigure}{1.0\linewidth}
            \begin{subfigure}{0.31\linewidth}\centering\scriptsize Seq.\end{subfigure}
            \begin{subfigure}{0.31\linewidth}\centering\scriptsize Base.\end{subfigure}
            \begin{subfigure}{0.31\linewidth}\centering\scriptsize Ours\end{subfigure}
        \end{subfigure}
    \end{subfigure}
	\caption{Comparison results at noise level $4$ (Gain $\propto 8$) on sample CLIC images for both MSE and MS-SSIM models. Apart from the better PSNR values and lower bpp rates, we can observe that the text and edges are better restored for our method.}
	\label{fig:qualitatives_clic}
\end{figure}

\begin{figure}[t!]
    \begin{subfigure}{1.0\linewidth}
        \begin{subfigure}{0.2\linewidth}
			\includegraphics[width=0.98\linewidth]{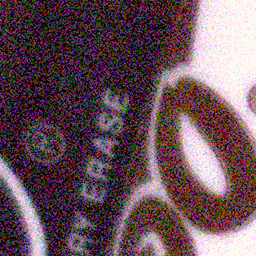}
    	\end{subfigure}%
        \begin{subfigure}{0.2\linewidth}
			\includegraphics[width=0.98\linewidth]{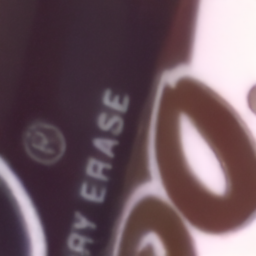}
    	\end{subfigure}%
        \begin{subfigure}{0.2\linewidth}
			\includegraphics[width=0.98\linewidth]{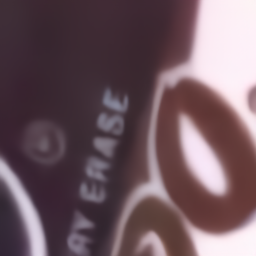}
    	\end{subfigure}%
        \begin{subfigure}{0.2\linewidth}
			\includegraphics[width=0.98\linewidth]{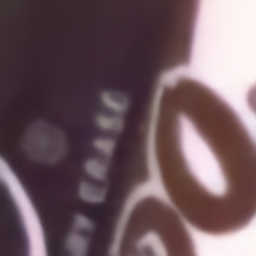}
    	\end{subfigure}%
        \begin{subfigure}{0.2\linewidth}
			\includegraphics[width=0.98\linewidth]{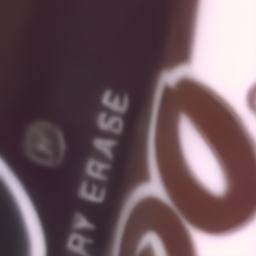}
    	\end{subfigure}%
	\end{subfigure}
	\\[2pt]
    \begin{subfigure}{1.0\linewidth}
        \begin{subfigure}{0.19\linewidth}\centering\scriptsize Noisy\end{subfigure}
        \begin{subfigure}{0.19\linewidth}\centering\scriptsize DeamNet\\(24.000bpp)\end{subfigure}
        \begin{subfigure}{0.19\linewidth}\centering\scriptsize Sequential\\(0.0459bpp)\end{subfigure}
        \begin{subfigure}{0.19\linewidth}\centering\scriptsize Baseline\\(0.0493bpp)\end{subfigure}
        \begin{subfigure}{0.19\linewidth}\centering\scriptsize Ours\\(0.0454bpp)\end{subfigure}
    \end{subfigure}
    \\[2pt]
    \begin{subfigure}{1.0\linewidth}
        \begin{subfigure}{0.2\linewidth}
			\includegraphics[width=0.98\linewidth]{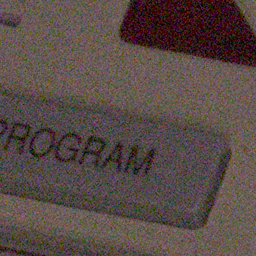}
    	\end{subfigure}%
        \begin{subfigure}{0.2\linewidth}
			\includegraphics[width=0.98\linewidth]{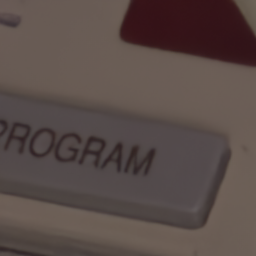}
    	\end{subfigure}%
        \begin{subfigure}{0.2\linewidth}
			\includegraphics[width=0.98\linewidth]{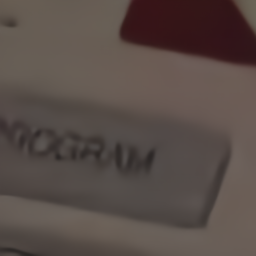}
    	\end{subfigure}%
        \begin{subfigure}{0.2\linewidth}
			\includegraphics[width=0.98\linewidth]{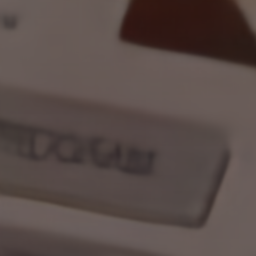}
    	\end{subfigure}%
        \begin{subfigure}{0.2\linewidth}
			\includegraphics[width=0.98\linewidth]{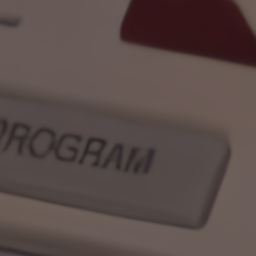}
    	\end{subfigure}%
	\end{subfigure}
	\\[2pt]
    \begin{subfigure}{1.0\linewidth}
        \begin{subfigure}{0.19\linewidth}\centering\scriptsize Noisy\end{subfigure}
        \begin{subfigure}{0.19\linewidth}\centering\scriptsize DeamNet\\(24.000bpp)\end{subfigure}
        \begin{subfigure}{0.19\linewidth}\centering\scriptsize Sequential\\(0.0234bpp)\end{subfigure}
        \begin{subfigure}{0.19\linewidth}\centering\scriptsize Baseline\\(0.0230bpp)\end{subfigure}
        \begin{subfigure}{0.19\linewidth}\centering\scriptsize Ours\\(0.0205bpp)\end{subfigure}
    \end{subfigure}
	\caption{Sample results on the SIDD. Since no ground-truth image is available for SIDD benchmark dataset, the visual results of DeamNet is shown as a reference for ground truth. We can see that the texts in our results is clearer at even slightly lower bpp rate.}
	\label{fig:qualitatives_sidd}
\end{figure}

\subsection{Efficiency Performance}
\label{speed}
We also compare the efficiency between our method and the sequential solution on the Kodak dataset, where the main difference comes from the compression process. The average elapsed encoding time under all qualities and noise levels for the sequential method is $75.323$ seconds, while our joint solution is only $7.948$ seconds. The elapsed running time is evaluated on Ubuntu using a single thread on Intel(R) Xeon(R) Gold $5118$ CPU with $2.30$GHz frequency. The sequential method has considerably longer running time than our joint method, where the additional overhead mainly comes from the heavy individual denoising modules in the encoding process of the sequential method. On the contrary, our joint formulation with efficient plug-in feature denoising modules, which pose little burden upon running time, is more attractive in real-world applications.

\subsection{Qualitative Results}
Some qualitative comparisons are presented to further demonstrate the effectiveness of our method. We show the visual results at noise level $4$ (Gain $\propto 8$) of the sample Kodak images in Fig.~\ref{fig:qualitatives_kodak} and CLIC images in Fig.~\ref{fig:qualitatives_clic} for both MSE and MS-SSIM models. Fig.~\ref{fig:qualitatives_sidd} shows the results of two sample patches from the SIDD. These results show that our method can obtain better quality images with even lower bit rates. Please check our supplements for more visual results.

\section{Conclusion}
We propose to optimize image compression via joint learning with denoising, motivated by the observations that existing image compression methods suffer from allocating additional bits to store the undesired noise and thus have limited capacity to compress noisy images. We present a simple and efficient two-branch design with plug-in denoisers to explicitly eliminate noise during the compression process in feature space and learn a noise-free bit representation. Extensive experiments on both the synthetic and real-world data show that our approach outperforms all the baselines significantly in terms of visual and metrical results. We hope our work can inspire more interest from the community in optimizing the image compression algorithm via joint learning with denoising and other aspects.

\bibliographystyle{splncs04}
\bibliography{egbib}

\end{document}